\documentclass[final,5p,times,11pt,twocolumn,authoryear]{elsarticle}

\usepackage{amsmath,amsfonts,amssymb,amscd,amsthm,xspace}
\usepackage{bm}									

\usepackage{array}								
\usepackage{tabularx}							
\usepackage{booktabs}							
\usepackage{longtable}							
\usepackage{tabu}								
\usepackage{siunitx} 							
\usepackage{dcolumn}							
  \newcolumntype{d}[1]{D{.}{.}{#1}} 			
\usepackage[table]{xcolor}						
\usepackage{flafter} 							

\graphicspath{{./Figures/}}						
\usepackage{subcaption}							
\usepackage[figuresright]{rotating}				
\usepackage{lscape}								

\usepackage[hyphens]{url}		 				
\usepackage[hidelinks]{hyperref}				
\hypersetup{colorlinks=true, linkcolor=blue, urlcolor=blue, citecolor=blue, anchorcolor=blue}

\usepackage{enumitem}							
\usepackage{verbatim}							

\usepackage[UKenglish]{babel}					
\usepackage[UKenglish]{isodate}					
\cleanlookdateon								

\usepackage[linesnumbered,ruled,vlined]{algorithm2e}
\usepackage{algpseudocode}						

\AtBeginShipout{%
	\ifnum\value{page}>1 %
	\typeout{* Additional boxing of page `\thepage'}%
	\setbox\AtBeginShipoutBox=\hbox{\copy\AtBeginShipoutBox}%
	\fi
}

\newcommand{\fref}[1]{Figure~\ref{#1}}
\newcommand{\tref}[1]{Table~\ref{#1}}
\newcommand{\eref}[1]{Equation~\ref{#1}}
\newcommand{\cref}[1]{Chapter~\ref{#1}}
\newcommand{\sref}[1]{Section~\ref{#1}}

\newcommand{\alref}[1]{Algorithm~\ref{#1}}

\journal{Preprint submitted to arXiv}

\begin{document}

\begin{frontmatter}
\title{Long short-term memory networks and laglasso for bond yield forecasting: \\ Peeping inside the black box\tnoteref{t1}}

\tnotetext[t1]{Declarations of interest: none.}


\author[UoS-ECS]{Manuel Nunes\corref{correspondingAuthor}}
\ead{m.nunes@soton.ac.uk}

\author[UoS-ECS]{Enrico Gerding}
\ead{eg@ecs.soton.ac.uk}

\author[UoS-SBS]{Frank McGroarty}
\ead{f.j.mcgroarty@soton.ac.uk}

\author[UoS-ECS]{Mahesan Niranjan}
\ead{mn@ecs.soton.ac.uk}

\address[UoS-ECS]{School of Electronics and Computer Science, University of Southampton, University Road, Southampton SO17 1BJ, United Kingdom}
\address[UoS-SBS]{Southampton Business School, University of Southampton, University Road, Southampton SO17 1BJ, United Kingdom}

\cortext[correspondingAuthor]{Corresponding author.}


\begin{abstract}

\noindent
Modern decision-making in fixed income asset management benefits from intelligent systems, 
which involve the use of state-of-the-art machine learning models and appropriate methodologies.
%
%
We conduct the first study of bond yield forecasting using long short-term memory (LSTM) networks, 
validating its potential and 
identifying its memory advantage.
%
Specifically, we model the 10-year bond yield using univariate LSTMs with three input sequences and five forecasting horizons. 
We compare those with multilayer perceptrons (MLP), univariate and with the most relevant features.
%
To demystify the notion of black box associated with LSTMs, 
we conduct the first internal study of the model.
%
To this end, we calculate the LSTM signals through time, at selected locations in the memory cell, 
using sequence-to-sequence architectures, uni and multivariate.
%
We then proceed to explain the states' signals
using exogenous information, 
for what we develop the LSTM-LagLasso methodology.
%
%
The results show that the univariate LSTM model with additional memory is capable of achieving similar results as the multivariate MLP using macroeconomic and market information.
Furthermore, shorter forecasting horizons require smaller input sequences and vice-versa.
%
The most remarkable property found consistently in the LSTM signals, 
is the activation/deactivation of units through time,
and the specialisation of units by yield range or feature.
%
Those signals are complex but can be explained by exogenous variables.
Additionally, some of the relevant features identified via LSTM-LagLasso are not commonly used in forecasting models.
%
%
In conclusion, our work 
validates the potential of LSTMs and methodologies for bonds,
providing additional tools for financial practitioners.

\end{abstract}


\end{frontmatter}

\section{Introduction}
\label{section-Introduction}

\noindent
The bond market, in particular the government sector, plays a fundamental role in the overall functioning of the economy and is of paramount importance for financial markets. This is the case, both as an asset class by itself (with an overall size of USD 102.0 trillion, as of 31-Dec-2016 \citep{bloomberg2017blgTerminal}, which compares to a global equity market of USD 66.3 trillion), and because the valuation methods of other asset classes often depend on bond yields as input information, especially for equities and corporate bond yields.
In addition, its importance derives from the fact that bonds and fixed income securities in general are a significant component of the portfolios of pension funds and insurance companies. 
Most commonly, the percentage of bonds varies from 25 to 40\% of portfolio assets for pension funds and around 40 to 70\% in the case of life insurance companies \citep{oecd2015outlook}.

Moreover, the bond market is at the early stages of both quantitative investing and electronic market-making, 
clearly lagging behind equities and foreign exchange (forex) markets.
%
%
%
%
%
This lag becomes evident in the scientific literature.
In fact, considerable attention has been devoted to the use of machine learning and development of techniques
for equity markets
%
\citep{booth2014automated, 							
	   eilersDunis2014tradingSeasonalEffects, 		
	   ballings2015stockPriceEnsembles, 			
	   dunis2016AIinFinancialMktsBook, 				
	   fischer2017lstmSP500, 						
	   krauss2017financialDisclosuresMiningLSTM,	
	   qin2017dualAttentionRNN, 					
       sermpinisKarathanasop2019stocksNNtrading},	
and also for forex markets
\citep{gradojevic2006FXforecasting, 				
	   huang2007NNinFXreview, 						
	   choudhry2012high, 							
	   sermpinisLawsKaraDunis2012FXwithNN,			
	   fletcher2013multiple,						
       sermpinis2013FXwithRBFand},					
just to mention a few studies (more detail in \sref{section-Literature-review}). 
In contrast, there is a significant gap in both the academic literature and the finance industry when it comes to the application of machine learning techniques in fixed income markets 
\citep{castellani2006forecasting, 					
	   dunis2007economic, 							
	   kanevski2008interest, 						
	   kanevski2010machine, 						
	   sambasivan2017GaussianProcessYC}.			

Furthermore, most of the applications of machine learning to financial assets
tend to be limited to forecasting and comparison of results versus benchmarks.
Only very few publications can be found that try to extract additional information from the model
or study how the model works
(see \sref{section-Literature-review}).
Indeed, advances in machine learning enable enhanced decision-making 
by e.g. using new types of data 
\citep{krauss2017financialDisclosuresMiningLSTM} 
and reinforcement learning techniques 
\citep{eilersDunis2014tradingSeasonalEffects}. 
However, for machine learning models to be useful in asset management decision-making they need to be trustworthy. 
To achieve this, a better understanding of their functioning is crucial.
%

%
In this field of machine learning, more precisely in deep learning,
one of the most successful models for sequence learning is the long short-term memory (LSTM) networks. 
The architecture of this model includes a feedback loop mechanism that enables the model to ``remember" past information.
This model has been achieving top results in other scientific fields but has not been used on a broad basis in financial applications.
This will be further detailed in \sref{section-Literature-review}.
More specifically, in the case of bonds, they have not been studied previously with LSTMs. 
This is an additional gap in the literature, 
despite both the importance of this asset class in financial markets and 
the potential of LSTMs for financial forecasting. 
The discussion of its potential will be the focus of 
\sref{section-DeepLearningModel-RNN-LSTM-advantages-limitations-potential}.

Given the status quo on machine learning research in bonds,
the main high-level objectives of this study are twofold:
to assess the potential of LSTM networks for bond yield forecasting,
testing their memory advantage versus memory-free models such as standard feedforward neural networks; and
to demystify the preconceived notion of black box associated to the LSTM model.
Together these objectives go towards bridging the gaps identified in the literature and presented above.
Besides, they contribute to improved knowledge and trustworthiness of LSTM networks, 
providing asset management practitioners with 
additional tools for better decision-making.

In more detail, our key contributions are as follows.
First,
we conduct an innovative application of a deep learning model (LSTM) to bonds.
The results are compared to memory-free multilayer perceptrons (MLP).
Our results validate the potential of LSTM networks for yield forecasting. This enables their use in intelligent systems for the asset management industry, in order to support the decision-making process associated with the activities of bond portfolio management and trading.
Additionally, we identify the LSTM's memory advantage over standard feedforward neural networks, showing that the univariate LSTM model with additional memory is capable of achieving similar results as the multivariate MLP with additional information from markets and the economy.

Second,
we go beyond the application of LSTMs,
by conducting an in-depth study of the model itself (opening the black box),
to understand the representations learned by its internal states.
Such explanations of what black-box models learn is a popular topic of interest for certification and litigation purposes.
In more detail, we extract and analyse the signals in both states (hidden and cell) and at the gates
inside the LSTM memory cell. 
This is the first contribution to demystify the notion of black box attached to LSTMs
using a technique which is fundamentally different from the most relevant ones found in the literature. 
Other studies are applied to a different type of recurrent neural network 
\citep{giles2001TSpredictionRNN}
or perform an external analysis of the model 
\citep{fischer2017lstmSP500}.

Third and last, 
following the extraction of signals at those locations,
we proceed to explain the information they contain with exogenous economic and market variables.
For that purpose, we develop a new methodology here identified as LSTM-LagLasso, based on both
Lasso \citep{tibshirani1996regression} and 
LagLasso \citep{mahler2009modeling}.
This methodology is capable of identifying both relevant features and corresponding lags.

The remainder of this paper is structured as follows. 
In \sref{section-Literature-review} the literature review is presented. 
In \sref{section-Deep-learning-model} 
we introduce the theory behind the deep learning model used in our research, together with its main advantages, limitations and potential for yield forecasting.
\sref{section-LSTM-networks-for-bond-yield-forecasting}
covers the bond yield forecasting study using LSTMs versus MLPs.
\sref{section-Opening-the-LSTM-black-box} 
focuses on the internal analysis of signals inside the LSTM model, 
while \sref{section-New-LSTM-LagLasso-method} 
details the explanation of those signals using exogenous variables, introducing the LSTM-LagLasso methodology developed for that purpose.
Finally, in \sref{section-Conclusions-and-future-work}
the main conclusions are outlined together with direction for future work.

\section{Literature review} 
\label{section-Literature-review}


\noindent
The literature review starts by looking at the main applications of recurrent neural networks in finance and other fields.
Then, considering the subset of publications on forecasting financial assets,
we analyse in detail the whole scope of the application carried out,
to compare and contrast with our research.

\subsection{RNN-LSTM applications in finance and other fields}
\label{section-LiteratureReview-RNN-LSTM-applications-in-finance-and-other-fields}

\noindent
The recurrent neural network family of model, especially the LSTM networks, has shown to significantly outperform.
In fact, in one of the most recent and comprehensive books on deep learning the authors categorically state that gated RNNs are the most effective sequence models used in practical applications, i.e. LSTMs and gated recurrent unit (GRU) based networks \citep{goodfellow2016deepLearningBook}.

The applications are almost endless, and comprise a wide variety of activities and scientific fields, such as 
(see \citet{lipton2015RNNreview}): 
handwriting recognition, text generation, natural language processing (recognition, understanding and generation), time series prediction, video analysis, musical information retrieval, image captioning, music generation, and in interactive type of problems like controlling a robot. For natural language processing in particular, LSTMs are among the most widely used deep learning models to date.

Most of these activities have in common the fact that they have sequential data. And this is what RNNs and LSTMs do best. They can process sequences as input, as output or in the most general case on both sides \citep{karpathy2015unreasonableRNN}. Furthermore, the LSTMs are used to take advantage of their capability to learn long-term dependencies.

In the financial domain, 
no publication could be found in the current literature using this type of model in fixed income markets. A discussion of its potential in this area will be the focus of \sref{section-DeepLearningModel-RNN-LSTM-advantages-limitations-potential}. 
Indeed, applications of RNN and LSTM models were found first and foremost for equities. See, for example, the works by \citet{xiong2015googleStanfordPaperStocks}, 		
\citet{diPersio2016lstmStocks, 						
	   diPersio2017lstmStocksGoogle}, 				
\citet{fischer2017lstmSP500}, 						
\citet{krauss2017financialDisclosuresMiningLSTM},	
\citet{munkhdalai2017lstmStocks} and 				
\citet{qin2017dualAttentionRNN}. 					
In addition, substantial work can also be found on forex markets 
\citep{giles2001TSpredictionRNN, 					
	   maknickiene2012lstmFX, 						
	   diPersio2016lstmStocksFX}. 					
Other applications can be detected for financial crises prediction 
\citep{gilardoni2017lstmFinancialDistress} 
and for credit risk evaluation in P2P platforms, or peer-to-peer lending 
\citep{zhang2017lstmCreditRiskP2P}.

\subsection{Main scope in the financial applications}
\label{section-LiteratureReview-Main-scope-in-the-financial-applications}

\noindent
Considering the applications to other financial assets presented in 
\sref{section-LiteratureReview-RNN-LSTM-applications-in-finance-and-other-fields},
what most of them have in common is that they are pure applications of the model.
This usually includes 
the implementation of the model to the selected asset class (equity, forex, or other),
the calculation of errors, and
the comparison with other models or benchmarks.
%
%
In some notable exceptions, the research goes beyond the pure application, and 
analyses the information inside the model.
The main studies in this context are presented below.

To begin with, \citet{strobelt2018lstmvis} have developed a tool that facilitates the visualisation of signals inside the LSTM memory cell, called ``LSTMVis".
This analysis can be used to identify hidden state dynamics in the LSTM model that would otherwise be lost information. 
There are some indications in the literature that this type of information extraction from the gates may be relevant, as will be seen below. 
%
This visualisation tool is especially orientated to the manipulation of sequences of words.
They may consist of English words for translation or sentiment detection, or
other type of symbolic input as musical notes or code.

Based on the ``LSTMVis" visualisation tool,
\citet{diPersio2017lstmStocksGoogle} provide examples of activation time series in RRNs.
The authors mention that this type of analysis may be able to detect trends in time series and,
in this case, the signals could be used as indicators.
%
However, it is not specified in their paper at what level in the model the authors captured those signals (states or gates and their identification).
Nevertheless, this study represents a first attempt at this subject, although further research is needed 
to demonstrate the mentioned detection of trends capability in time series.

In an earlier study, \citet{giles2001TSpredictionRNN} refer to the possibility of extracting rules and knowledge from trained recurrent networks modelling noisy time series. 
To that end, they first convert the input time series into a symbolic representation using self-organizing maps. 
Then the problem becomes one of grammatical inference and they use RNNs considering the sequence of symbols as inputs. 
More specifically, they use an Elman type of architecture for the recurrent neural network \citep{elman1990RNNinitial}.
In addition, the converted inputs facilitate the extraction of symbolic information from the trained RNNs
in the form of deterministic finite state automata.
%
The interpretation of that information resulted in the extraction of simple rules, 
such as trend following and mean reversion.

In contrast to \citet{giles2001TSpredictionRNN},
in our research we use the LSTM model, a more recent type of recurrent neural networks.
Moreover, it should be emphasized that the conversion into symbols has a filtering effect on the data
and this may be undesirable.
Hence, our data pre-processing does not include any type of filtering operation given our view that the data complexity and its volatility do not conform with the concept of noise. 
This issue will be further detailed in
\sref{section-NewLSTMLagLassoMethod-Methodology} -- LSTM-LagLasso.

Last but not least, 
\citet{fischer2017lstmSP500} have used a different approach
with the same objective of interpreting what happens in the model.
%
We will briefly explain the methodology used, for context and to clarify the differences in relation to our approach.
In their research, the authors carry out a notable and comprehensive market prediction study on the component stocks of the S\&P 500.
Using LSTMs they forecast the next time step, subsequently ranking the individual stocks based on the probability of outperforming the cross-sectional median.
Then they group the $ k $ top and bottom stock performers ($k \in \{10, 50, 100, 150, 200\}$).
Considering the model's input sequence of returns for each of those groups,
they calculate several descriptive statistics
and identify characteristics of the stocks belonging to each of those two groups (top and bottom likely performers).
%
%
Using this methodology, they found that stocks in the top and bottom group exhibit the following characteristics:
high volatility, below-mean momentum, and extreme returns in the last few days with a tendency to revert them in the short-term.
This is especially the case for groups with a smaller number of stocks (smaller $ k $).
Since these characteristics were found on direct outputs of the model, 
they are attributed to the functioning of the LSTM networks.

Based on the result obtained, 
\citet{fischer2017lstmSP500} devised a trading strategy, which consists of 
selling the (recent past) winners and buying the (recent past) losers.
%
This is a possible but simplified trading strategy.
It is known in practice in financial markets as ``contrarian", 
in the sense that it is counter-intuitive, and
it is certainly far from consensual.
See for example the work of 
\citet{jegadeesh1993tradingBwinnersSlosers} and 
\citet{wang2019RELEARNtradingBwinnersSlosers}
supporting the opposite strategy ``buy the winners and sell the losers".
On the contrary, 
\citet{khanal2014tradingBwinnersSlosersNoEvidence}
found no clear evidence of a profitable ``buy the winners and sell the losers" strategy, 
considering it on a buy and hold basis for 3 to 12-month periods and 
for the period studied between 1990 and 2012.
And finally, the work of 
\citet{antoniou2003tradingBlosersSwinners} 
supporting the strategy
``buy the losers and sell the winners", considered in the work in analysis.
In fact, in the most dangerous situations for a stock and most penalising for a portfolio return, a sequence of negative returns may be just the beginning of a serious bear market for that particular stock, or something even more serious affecting the company that will result in a prolonged correction.

To conclude, the authors \citep{fischer2017lstmSP500} conducted an external analysis of the LSTM model, in the sense that they use the output of the model to infer their functioning. No internal analysis is carried out of the LSTM states or gates, and this is the main difference in relation to our research and one of our contributions to the present state-of-the-art.

\section{Deep learning model}
\label{section-Deep-learning-model}


\noindent
In this section, we present the deep learning model used for this study, namely the LSTM networks. The other model considered is not described here. Further information on standard feed-forward neural networks / MLP can be found elsewhere (see, for example, \citet{bishop2006machineLearningBook}, \citet{hastie2013statisticalLearningBook} and \citet{rumelhart1986learning}, the latter for the training process using the back-propagation algorithm). 
Then we discuss the potential of LSTMs for yield forecasting.

\subsection{Long short-term memory networks}
\label{section-DeepLearningModel-Long-short-term-memory-networks}

\begin{figure*}[!htb]
	\centering
	\captionsetup{width=1.00\textwidth}
	\includegraphics[width=0.85\textwidth]{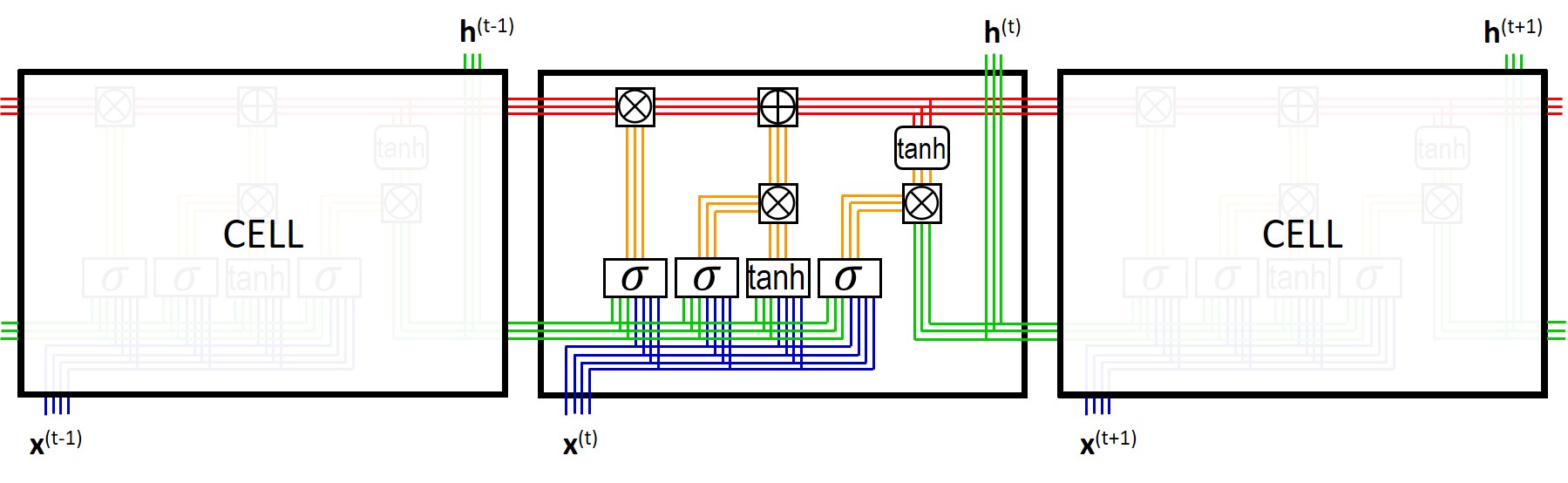}
	\caption{Long short-term memory cells unrolled in time (based on \citep{olah2015understandingLSTM}, with modifications made by the authors).}
	\label{Figure-LSTM-cells}
\end{figure*}

\begin{figure*}[!htbp]
	\centering
	\captionsetup{width=1.00\textwidth}
	\includegraphics[width=0.85\textwidth]{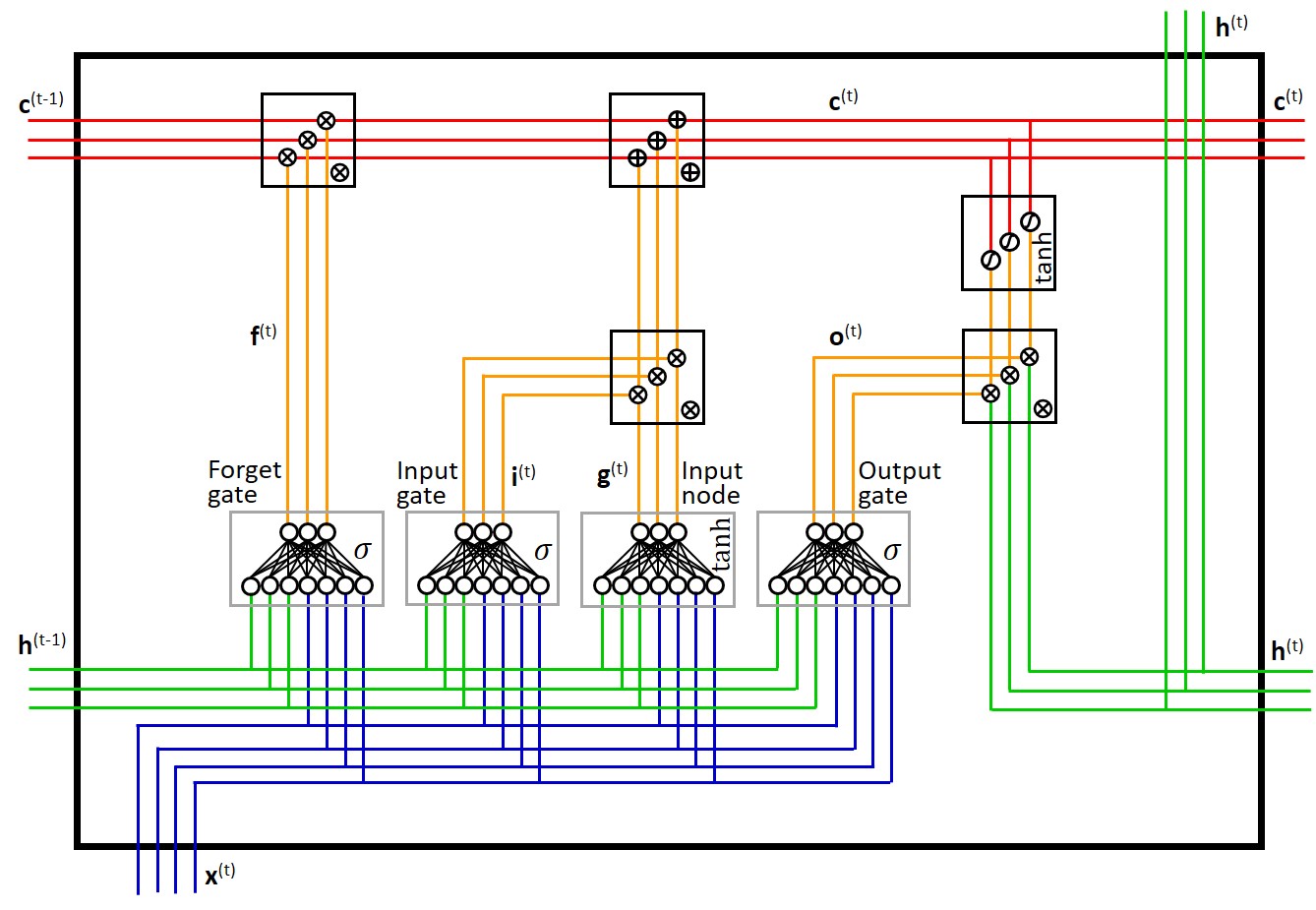}
	\caption{Long short-term memory detailed cell diagram (based on \citep{prugelBennett2017advMachineLearning}, with modifications made by the authors).}
	\label{Figure-LSTM-main-diagram}
\end{figure*}

\noindent
The LSTM architecture was first introduced by \citet{hochreiterSchmidhuber1997LSTMinitial}, and subsequently adapted by other researchers \citep{gers1999lstmLearningToforget, gersSchmidhuber2000lstmPeepholeConnections, gravesSchmidhuber2005lstmPlus}.
The LSTM model is a type of recurrent neural network, having a structure that includes a clever feedback loop mechanism delayed in time, and this structure can be ``unrolled" in time.
At each time step, the LSTM cell has a structure which is substantially more complex than a standard RNN, incorporating four complete neural networks in each of those cells (also called memory cells). 
In \fref{Figure-LSTM-cells} a simplified diagram of the unrolled chain-type structure is presented, identifying the main components of a memory cell.
A more detailed representation of an LSTM cell is presented in \fref{Figure-LSTM-main-diagram}.

The corresponding equations that govern the modern LSTM model can be expressed in the following form, with \fref{Figure-LSTM-main-diagram} showing where these operations occur in the LSTM cell \citep{hochreiterSchmidhuber1997LSTMinitial, lipton2015RNNreview, goodfellow2016deepLearningBook}:

%
\begin{eqnarray}
\bm{f}^{(t)} &=& \sigma ~ (\bm{W}^{fx} ~ \bm{x}^{(t)} + \bm{W}^{fh} ~ \bm{h}^{(t-1)} + b_{f})  
\label{Equation-LSTM-01}   \\
\bm{i}^{(t)} &=& \sigma ~ (\bm{W}^{ix} ~ \bm{x}^{(t)} + \bm{W}^{ih} ~ \bm{h}^{(t-1)} + b_{i})  
\label{Equation-LSTM-02}   \\
\bm{g}^{(t)} &=& tanh ~   (\bm{W}^{gx} ~ \bm{x}^{(t)} + \bm{W}^{gh} ~ \bm{h}^{(t-1)} + b_{g})  
\label{Equation-LSTM-03}   \\
\bm{o}^{(t)} &=& \sigma ~ (\bm{W}^{ox} ~ \bm{x}^{(t)} + \bm{W}^{oh} ~ \bm{h}^{(t-1)} + b_{o})  
\label{Equation-LSTM-04}    
\end{eqnarray}
\begin{eqnarray}
\bm{c}^{(t)} &=& \bm{f}^{(t)} \otimes ~ \bm{c}^{(t-1)} + \bm{i}^{(t)} \otimes ~ \bm{g}^{(t)}
\label{Equation-LSTM-05}   \\
\bm{h}^{(t)} &=& \bm{o}^{(t)} \otimes ~ tanh(\bm{c}^{(t)})
\label{Equation-LSTM-06}
\end{eqnarray}

\noindent
where 
$ \bm{f}^{(t)} $ 						is the function for the forget gate;
$ \bm{i}^{(t)} ~and~ \bm{g}^{(t)} $ 	are the functions for the input gate and for the input node, respectively;
$ \bm{o}^{(t)} $ 						the function for the output gate;
$ \bm{c}^{(t)} ~and~ \bm{c}^{(t-1)} $	is the cell state (also called internal state) at time step $t$ and $t-1$;
$ \bm{h}^{(t)} ~and~ \bm{h}^{(t-1)} $	the hidden state at time step $t$ and $t-1$;
$ \bm{x}^{(t)} $						is the input vector at time step $t$;
$ W $									are the weight matrices, with 
$ \bm{W}^{fx} $							as an example, representing the weight matrices for the connection input-to-forget gate 
(indices indicating to-from connections);
$ \bm{b}_{f}, \bm{b}_{i}, \bm{b}_{g}, \bm{b}_{o} $	are the bias vectors;
$ \sigma $								is the logistic sigmoid activation function;
$ tanh $ 								the hyperbolic tangent activation function; and
$ \otimes $								represents the Hadamard product (i.e. element-wise multiplication).


\subsection{LSTM advantages, limitations and potential for yield forecasting}
\label{section-DeepLearningModel-RNN-LSTM-advantages-limitations-potential}

\noindent
After describing the architecture of the LSTMs we now identify the advantages, limitations and potential for yield forecasting. 
%
First, the main advantage of the LSTM model is related to the reason why it was developed in the first place. The RNNs could not capture the long-term dependencies due to the vanishing or exploding gradients problem 
first identified by \citet{hochreiter1991RNNvanishingExplodingGradientsInitial} in a diploma thesis
\citep{hochreiter2001RNNvanishingExplodingGradientsEnglish, schmidhuber2015deepLearningOverview},
and in parallel research work by 
\citet{bengio1993RNNvanishingExplodingGradientsInitial2, bengio1994RNNvanishingExplodingGradientsInitial2}.
In fact, this memory capability is one of the characteristics that clearly separates this type of model from the standard neural networks, in particular the MLPs. Given the stateless condition of MLP models they only learn fixed function approximations.

For modelling financial time series, it seems probable that long-term dependencies are important. Even though the last available value of the series is the one collecting all the available information in the market up to the present moment, inversion points tend to follow certain patterns, frequently exploited by technical analysts who look essentially at chart data. A model with memory and capable of learning long-term dependencies may be beneficial for this reason.

Second, in sequence prediction problems, the sequence imposes an order on the data that must be respected when training and forecasting, i.e. the order of the observations is important for the modelling process. This is the case with financial time series. However, in feed-forward neural networks / MLPs, the modelling of time series' temporal structure is only done indirectly through the consideration of multiple time steps as different input features. Although with this method previous values are included in the regression problem, the natural ``sequence" or structure of the time series is not really present in the modelling process and the model does not have any knowledge of it. 
The LSTMs are the most effective models for sequence learning, modelling these time sequences directly. 
Additionally, it can input and output sequences time step by time step, enabling variable length inputs and/or outputs. With this property, they overcome one of the main limitations of standard feedforward neural networks.

Third and last, a model for financial time series should be able to perform multi-asset forecasting, for the prediction of several targets simultaneously and it would be desirable to perform multi-step forecasting, to consider several forecasting horizons into the future.
The LSTM type of model is capable of dealing with multivariate problems and also with multi-step prediction using sequence-to-sequence architectures, thus fulfilling these requisites naturally.

On the limitation side, some time series forecasting problems are technically simpler, not requiring the characteristics of a recurrent type of model. This is the case in particular when the most relevant data for making the prediction is within a small window of recent historic values. Here, the capability to deal with long-term dependencies and the model ``memory" are clearly not necessary. In this type of situation, MLPs and even linear models may outperform the LSTM pure-autoregressive univariate model, with lower complexity 
\citep{gers2002lstmTimeWindow, brownlee2018bookLSTM}.

Overall, given the additional complexity of the model, it should only be used when the type of problem we have is better modelled by this type of neural network architecture. And this is reflected in two main conditions: sequential data and when the long-term dependencies may help the forecasting process. Nonetheless, LSTM networks' potential for yield forecasting seems evident. Despite being predominantly used for non-financial applications, their characteristics make them potentially suitable for financial time series predictions.

\section{LSTM networks for bond yield forecasting}
\label{section-LSTM-networks-for-bond-yield-forecasting}


\noindent
Now that we have discussed the LSTM model and its potential for yield forecasting, we move on to the empirical work carried out.
In this section, we describe the choices made on the dataset used, identifying the target, features and pre-modelling operations (including the generation of additional features for the MLP model, the train-test split and normalisation).

\subsection{Data}
\label{section-MemoryAdvantageOfLSTMs-Data}


\noindent
Given the interconnectedness and mutual influence of various asset classes in the markets, we consider a large number of features from financial markets. These are selected from government bond markets and from related classes and indicators: credit (corporate bonds), equities, currencies, commodities and volatility. 
Additional features are added, which are calculated from the previously mentioned features, specifically, bond spreads, slope of the yield curve and simple technical analysis indicators. 
Furthermore, economic variables are also very important, as clearly exemplified by the well established yields-macro models such as the Dynamic Nelson--Siegel model \citep{diebold2006DynamicNelsonSiegelmodel}. Hence, a vast range of economic indicators are also included, from different geographic locations. 
The complete list includes 159 features and, because it is so extensive, it is stored and made 
publicly available \citep{nunes2019dssPaperDataset}.
The target chosen for this study is the generic rate of the 10-year Euro government bond yield.

The dataset was obtained from Bloomberg database \citep{bloomberg2017blgTerminal} and covers the period from January 1999 to April 2017, giving 4779 time points of data. 
The former is the starting date for most time series of the Euro benchmarks and the total period covers several bull and bear markets.
Regarding data frequency, the selection was daily closing values, which are easily available for financial assets in general.

As mentioned previously in \sref{section-DeepLearningModel-RNN-LSTM-advantages-limitations-potential}, in the case of feed-forward neural networks / MLPs, the modelling of time series' temporal structure is done indirectly through the consideration of multiple time steps as different input features. Contrary to the MLP work, there is no need to generate additional features for the LSTM network, since previous time steps are given directly to the model in the form of an input sequence.
Hence, for the MLPs, new features are generated from the original ones, corresponding to lagged values of the respective time series. In our research, six time steps are considered, based on previous studies \citep{mahler2009modeling}. 
%
The selection of the most relevant features is carried out using Lasso (standing for Least Absolute Shrinkage and Selection Operator) regression proposed by \citet{tibshirani1996regression}.

As is common, we divide the data into two groups, for training and testing the models. In this case, a 70\% / 30\% split is considered. 
This refers to the overall static training and testing datasets. 
The static training dataset is considered for tuning hyperparameters, by sub-dividing it into new training and validations datasets, 
ensuring that results are quoted on totally unseen (test) portions of data.
When forecasting time series points from the static testing dataset, the actual training data is a dynamic moving window spanning up to the present moment for the time step being considered.

Finally, all data is normalised by subtracting the mean and dividing by the standard deviation of the training dataset (the dynamic moving window). This is also essential, given the wide range of features we are considering, which have very different scales in some cases.
For example, the 10-year yield is quoted in percentage, usually staying below 6\%, while several equity indices reach values well above 10000 (Dow Jones, Nikkei 225 and Hang Seng).

\subsection{Methodology}
\label{section-MemoryAdvantageOfLSTMs-Methodology}






\noindent
In this section, the empirical work carried out is identified and explained:
first, we compare directly the univariate MLP and LSTM models;
then, we further assess the LSTM potential for yield forecasting using different input sequences. 
Moreover, the models considered are specified and additional aspects of the methodology adopted are described, in particular in what concerns moving windows, retraining of models and cross-validation. 
A summary of the models used and additional model information is presented in \tref{table-Summary-of-empirical-work-with-LSTMs-vs-MLPs}.

\begin{table}[!htb]
	\bigskip
	\centering
	\caption{Summary of the models used.}
	\label{table-Summary-of-empirical-work-with-LSTMs-vs-MLPs}
	
	\begin{tabularx}{1.0\columnwidth}{l l X}
		
		\toprule
		
		\multicolumn{3}{l}{\textbf{Model information}} \\
		\midrule
		
		\multicolumn{2}{l}{Original features}    & 159                                        \\
		\multicolumn{2}{l}{Generated features}   & 795                                        \\
		\multicolumn{2}{l}{Target}               & 10-year yield                              \\
		\multicolumn{2}{l}{Forecasting horizons} & 0 (next day), 5, 10, 15, 20                \\
		\multicolumn{2}{l}{Moving window size}   & 3000 days                                  \\
		
		\multicolumn{2}{l}{Hidden units MLP}     & 10                                         \\
		\multicolumn{2}{l}{Hidden units LSTM}    & 100                                        \\
		\multicolumn{2}{l}{Time steps MLP}       & 6 days                                     \\
		\multicolumn{2}{l}{Time steps LSTM}      & 6, 21, 61 days                             \\
		\midrule
		\textbf{Model} & \textbf{Short name}     & \textbf{Description}                       \\
		\midrule
		\multicolumn{3}{l}{\textit{Direct comparison MLP vs. LSTM}}                           \\
		\midrule
		
		MLP            & NN TgtOnly              & MLP with target data only                  \\
		LSTM           & LSTM06                  & LSTM using input sequence of 6 time steps  \\
		
		\midrule
		\multicolumn{3}{l}{\textit{LSTMs with different input sequences}}                     \\
		\midrule
		
		MLP            & NN RelFeat              & MLP with relevant features                 \\
		               & NN TgtOnly              & MLP with target data only                  \\
		
		\midrule
		
		LSTM           & LSTM06                  & LSTM using input sequence of 6 time steps  \\
		               & LSTM21                  & LSTM using input sequence of 21 time steps \\
		               & LSTM61                  & LSTM using input sequence of 61 time steps \\
		
		\bottomrule
		
	\end{tabularx}
	\bigskip
\end{table}

For a streamlined approach to the model, we apply LSTM networks to a univariate type of problem,
i.e. the model has only one feature, corresponding to 
past values of the target we want to predict.
This is justified by the fact that we want to be able to assess the LSTM model potential, so we prefer to make the comparison with MLPs in its most pure condition. 
In other words, we prefer to perform the comparison without introducing additional features in the LSTM model, 
which would introduce an extra level of complexity.

When forecasting for longer forecasting horizons beyond one step ahead, there are two methods that can be used: direct or iterative forecasting. 
On the one hand, in the direct forecast only current and past data is used to forecast directly the time step required, using a horizon-specific model. On the other hand, in the iterative forecast a one step ahead model is iterated forward until the target forecasting horizon is reached. 
All the models use direct forecasting of targets given that, with financial time series, the prediction errors tend to propagate fast if we were to make iterative predictions. As a result, new neural networks are trained for each forecasting horizon.

For the direct comparison between the univariate MLP and the univariate LSTM, 
an LSTM is selected that most closely simulates the conditions applied to the MLP neural network,
i.e. both considering 6 time steps, based on previous studies \citep{mahler2009modeling}. 
The corresponding models are (\tref{table-Summary-of-empirical-work-with-LSTMs-vs-MLPs}): 
Model ``NN TgtOnly", meaning neural network using only the target variable as feature; and
Model ``LSTM06".
%
%
To further assess the potential of LSTMs for yield forecasting, 
we extend the number of models on both sides.
On the LSTM side,  
we consider three different input sequences in total: 6, 21 and 61 time steps, 
with the last two corresponding to approximately 1 and 3 calendar months.
The selected number of time steps also follow the structure 
``next day" plus desired period, i.e. 1+20, 1+60,
which will be useful later on this work when using sequence-to-sequence LSTM architectures.
On the MLP side,
we add to the univariate model, 
the MLP using the most relevant features selected for the 10-year bond yield target and 
individually for each forecasting horizon (Model ``NN RelFeat").
Besides, the comparisons are carried out for all forecasting horizons considered, 
i.e. next day and (next day plus) 5, 10, 15 and 20 days ahead.

Both MLP and LSTM models are trained using moving windows and retraining of models at every time step. This technique is feasible in real time and is used to take full advantage of the models.
%

Throughout this study, the main metric used is the mean squared error (MSE), which is commonly used for this purpose.
The results in \sref{section-MemoryAdvantageOfLSTMs-Results-and-discussion} are presented in the normalised version, i.e. calculated directly from the real and predicted normalised yields (\sref{section-MemoryAdvantageOfLSTMs-Data}), since the non-normalised equivalents are scale dependent. Hence, the normalised metric is used to facilitate the comparison of models.

In terms of the number of hidden units for the LSTM, this is set to 100 units (\tref{table-Summary-of-empirical-work-with-LSTMs-vs-MLPs}). A higher number would require additional computational time and this is a good compromise between speed and accuracy.
This number is also compatible with the maximum number of relevant variables considered in the MLP study for modelling the longest forecasting horizons. 
Regarding the MLP number of hidden units, the main conclusion from the hyperparameter tuning is that 10 hidden units is a good compromise, with significant overfitting observed for neural networks with more than 100 units.

Finally, the optimiser chosen is Adam \citep{kingma2014adam}.
This is the algorithm for gradient-based optimization of the stochastic objective function, and is frequently used for this type of problem \citep{brownlee2018bookLSTM}.

\subsection{Results and discussion}
\label{section-MemoryAdvantageOfLSTMs-Results-and-discussion}


\noindent
In this section, the main results are presented for both studies: the direct comparison MLP vs. LSTM and the assessment of LSTM potential using different input sequences. This is followed by a discussion of its implications for the present research.

\begin{figure}[!htb]
	\centering
	\includegraphics[width=0.475\textwidth]{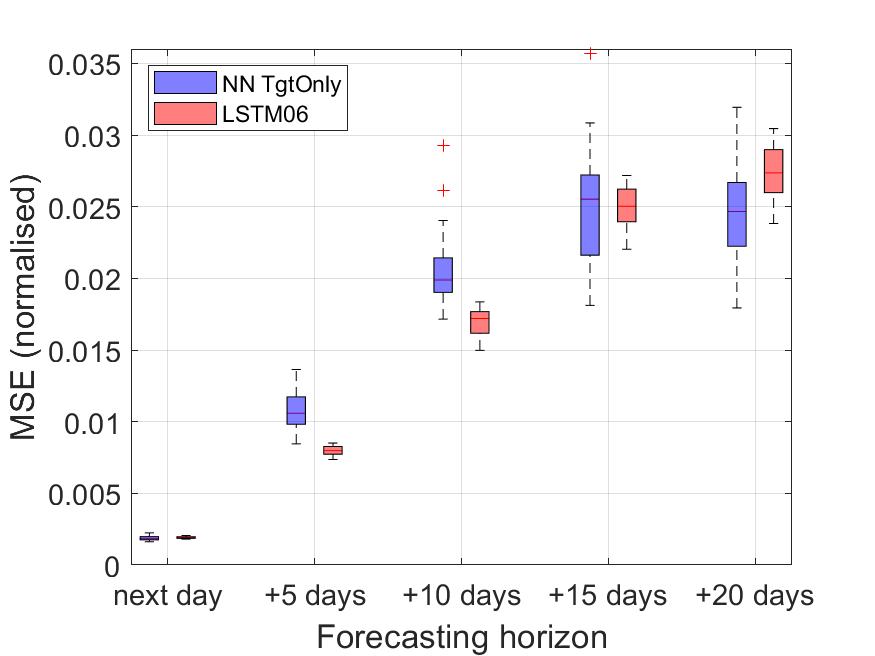}
	
	\captionsetup{width=0.475\textwidth}
	\caption{Direct comparison of models: univariate multilayer perceptron ``NN TgtOnly" vs. long short-term memory networks for input sequence of 6 time steps ``LSTM06".}
	\label{figure-LSTM-Comparative-Study-MLP-NNTgtOnly-vs-LSTM06}
\end{figure}

Starting with the first study,
the direct comparison of models, the results obtained are summarised in 
\fref{figure-LSTM-Comparative-Study-MLP-NNTgtOnly-vs-LSTM06}.
%
As can be seen, the results obtained with the LSTM model are better for forecasting horizon of 5 and 10 days,
achieving MSE reductions of 25\% and 14\%, respectively (median values), compared to the univariate MLP.
When the horizon is too large (20 days) the advantage of the LSTM is lost.
In all other cases, the results are not significantly different.

Another aspect that can be observed from the results is that the standard deviation obtained with LSTMs is lower for all horizons analysed. This is also an important outcome, giving indications of higher stability of this type of model when compared with the more traditional MLP model.



\begin{figure*}[!htbp]
	\centering
	\begin{subfigure}[t]{0.475\textwidth}
		\captionsetup{width=0.80\textwidth}
		\includegraphics[width=\textwidth]{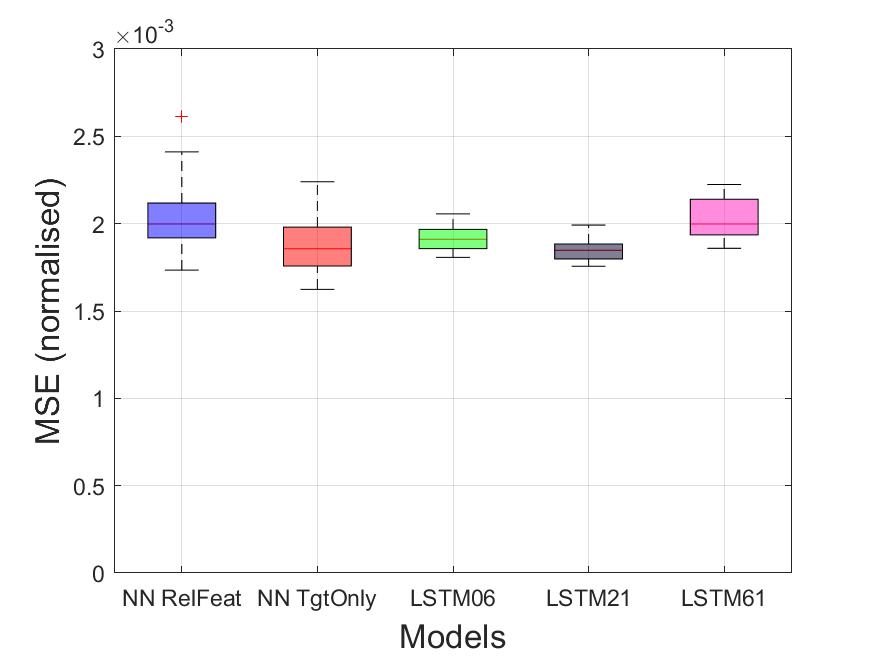}
		\caption{Forecasting horizon = 0 days (next day).}
		\label{figure-LSTM-Comparative-Study-cp00}
	\end{subfigure}\hfill%
	\begin{subfigure}[t]{0.475\textwidth}
		\includegraphics[width=\textwidth]{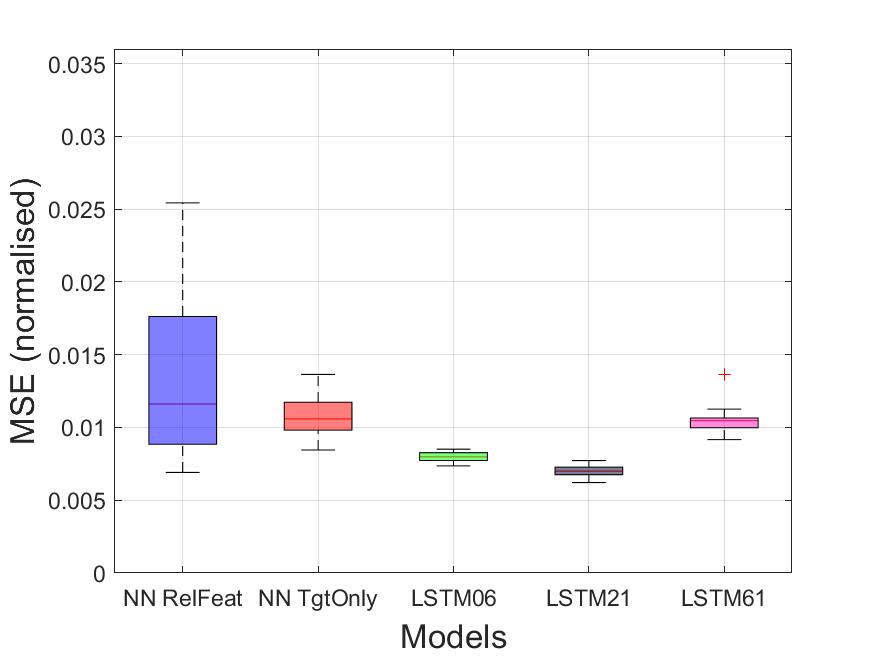}
		\caption{Forecasting horizon = 5 days.}
		\label{figure-LSTM-Comparative-Study-cp05}
	\end{subfigure}%
	
	\vspace{0.25cm}
	
	\begin{subfigure}[t]{0.475\textwidth}
		\includegraphics[width=\textwidth]{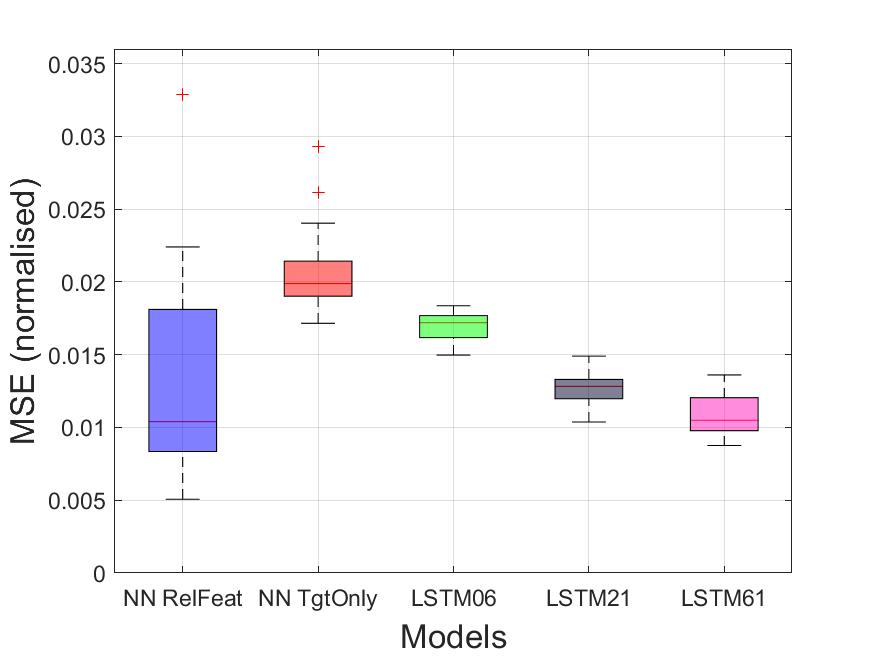}
		\caption{Forecasting horizon = 10 days.}
		\label{figure-LSTM-Comparative-Study-cp10}
	\end{subfigure}\hfill%
	\begin{subfigure}[t]{0.475\textwidth}
		\includegraphics[width=\textwidth]{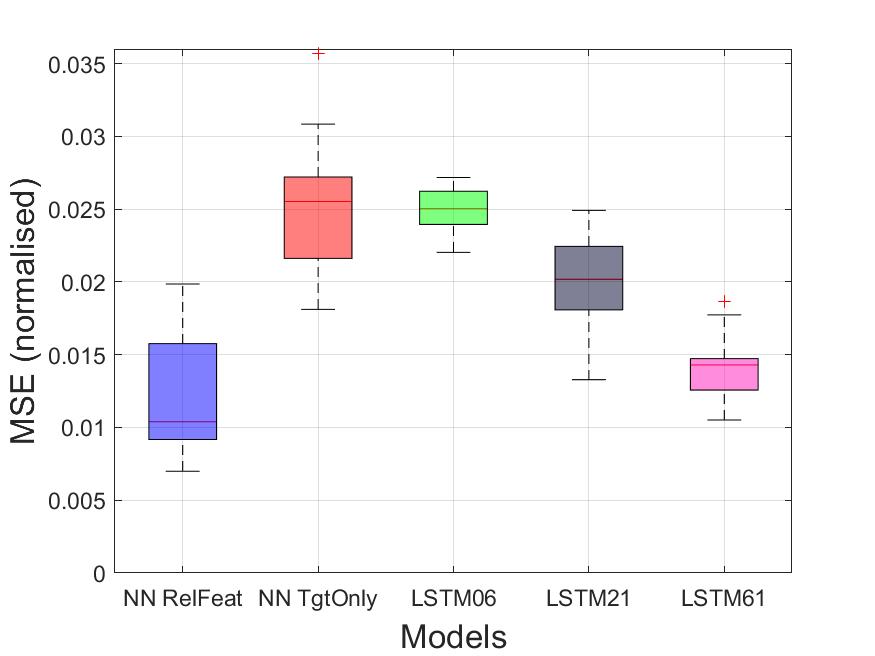}
		\caption{Forecasting horizon = 15 days.}
		\label{figure-LSTM-Comparative-Study-cp15}
	\end{subfigure}%
	
	\vspace{0.25cm}
	
	\begin{subfigure}[t]{0.475\textwidth}
		\includegraphics[width=\textwidth]{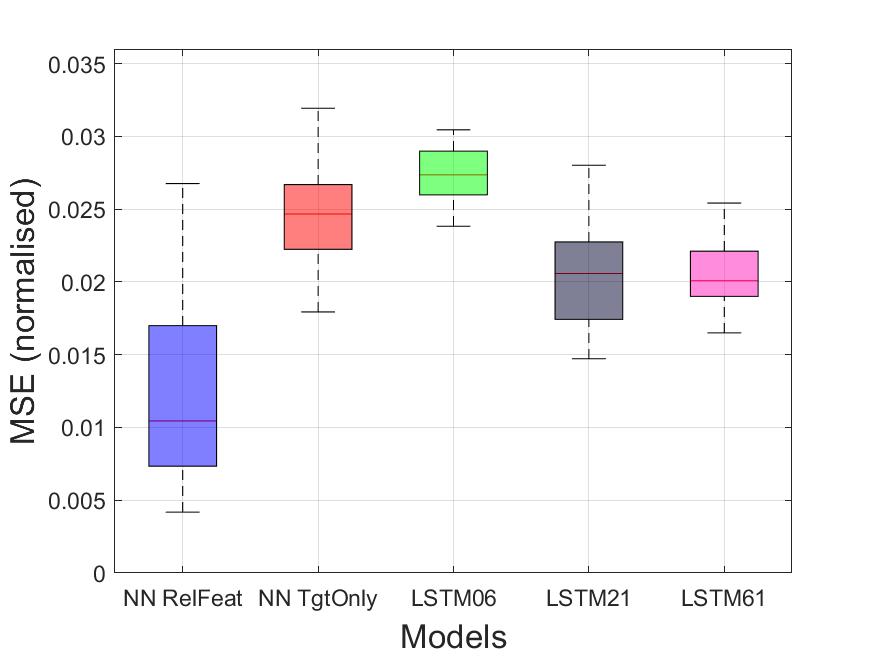}
		\caption{Forecasting horizon = 20 days.}
		\label{figure-LSTM-Comparative-Study-cp20}
	\end{subfigure}
	
	\vspace{0.25cm}
	
	\captionsetup{width=1.0\textwidth}
	\caption{Comparison of models: two types of multilayer perceptrons (MLP using the relevant features determined for the 10-year yield target and for each forecasting horizon ``NN RelFeat" and univariate MLP ``NN TgtOnly") vs. long short-term memory networks (LSTM) for input sequences of 6, 21 and 61 time steps (``LSTM06", ``LSTM21" and ``LSTM61", respectively).}
	
	\label{figure-LSTM-Comparative-Study-MLPs-vs-LSTM-all}
\end{figure*}

Regarding the second study,
LSTMs using different input sequences, the results are shown in 
\fref{figure-LSTM-Comparative-Study-MLPs-vs-LSTM-all}. 
%
When forecasting the next day (\fref{figure-LSTM-Comparative-Study-cp00}), the results obtained from all models are similar. This situation is equivalent to that reported in \citet{nunes2019eswaPaper} for next day forecasting with models including multivariate linear regression and a variety of MLP-type of models.
One striking advantage revealed by the LSTM is again the lower standard deviation in all LSTM models when compared to both MLPs.

However, it is when we forecast more distant time steps, that the benefits of LSTMs become more evident. 
Hence, considering the forecasting horizon of (next day plus) 5 days, as shown in \fref{figure-LSTM-Comparative-Study-cp05}, the LSTMs with input sequence length of 6 and 21 time steps (Models ``LSTM06" and ``LSTM21", respectively) produce results that significantly outperform both MLP models, with lower errors and much lower standard deviations. 
When compared to the univariate MLP, these models achieve MSE reductions of 25\% and 34\%, respectively (median values).
On the other hand, the LSTM with an input sequence of 61 days (Model ``LSTM61"), 
does not produce a reduction in error, 
generating similar median results also with significantly lower standard deviation. 
%
Therefore, it appears that forecasting 5 days ahead does not require such a long input sequence of 60 days.

When we consider forecasting horizons of 10 and 15 days (Figures \ref{figure-LSTM-Comparative-Study-cp10} and \ref{figure-LSTM-Comparative-Study-cp15}), all LSTMs tend to perform better than the univariate MLP model (Model ``NN TgtOnly") with lower standard deviations. 
The MSE reductions achieved range from 2\% to 47\%.
Another important observation is that the LSTMs with longer input sequences are able to reach similar levels of forecasting error to the MLP with the most relevant features (Model ``NN RelFeat"), again with lower standard deviation. 
In fact, the LSTM appears capable of compensating, at least partially, the lack of additional information from markets with additional memory via longer input sequences. 
This is a promising result for future work.

Finally, the differences are not so clear when we consider forecasting horizon equal to (next day plus) 20 days
(Figure \ref{figure-LSTM-Comparative-Study-cp20}). 
In this case, the LSTMs with longer input sequences (Models ``LSTM21" and ``LSTM61") perform better than the univariate MLP (MSE reductions of 17\% and 19\%, respectively), with slightly lower standard deviation. 
However, the shorter sequence LSTM (Model ``LSTM06") produces slightly worse median value (MSE increase of 11\%), although with lower standard deviation. 
A possible explanation for this may be that the input sequence length of 6 time steps is already insufficient for forecasting 20 days ahead. Thus, either additional features or longer input sequences are required.

These results suggest that the LSTM architecture needs to take into account the specific problem we are trying to solve and the type of forecasting horizon we aim to predict. 
Additionally, the results of these comparisons carried out over a wide range of input and forecasting horizons suggest that, under some conditions, the structure in the data is better captured by having models with time delays in them (i.e. LSTMs), which can strike a balance between the use of immediate and distant past data.
However, the advantage of one class of models over others is not universal, as indicated by the no free lunch theorem 
\citep{wolpert1997noFreeLunchTheorems}.
In the next section, we probe the LSTM further, opening the black box, to see if the representations learned in its internal states are interpretable in any way. Such explanations of what black-box models learn is a popular topic of interest for certification and litigation purposes.


\section{Opening the LSTM black box: signals analysis}
\label{section-Opening-the-LSTM-black-box}


\noindent
Neural network methods in general and the LSTM in particular are considered black box type of models.
They establish functional relationships between inputs and outputs, 
but one cannot extract interpretable information from the model itself.
In this section we present our contribution to demystify this complex deep learning model,
by analysing the signals inside the LSTM memory cell and extracting relevant information.
In turn, this is subsequently used to identify relevant explanatory features for this problem
(\sref{section-New-LSTM-LagLasso-method}).
This section also includes
a brief explanation of the LSTM memory cell, its states and gates,
since they are indispensable to understand the methodology used.

\subsection{Memory cell: states and gates}
\label{section-OpeningTheLSTMBlackBox-Memory-cell-states-and-gates}

\begin{figure*}[!htb]
	\centering
	\begin{subfigure}[t]{0.5\textwidth}
		\captionsetup{width=1.0\textwidth}
		\includegraphics[width=\textwidth]{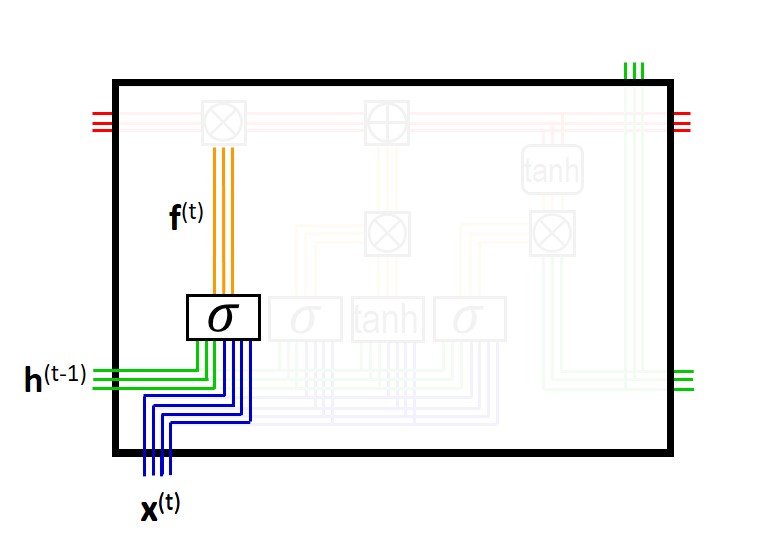}
		\caption{Forget gate.}
		\label{figure-LSTM-gates-forget-gate}
	\end{subfigure}\hfill%
	\begin{subfigure}[t]{0.5\textwidth}
		\includegraphics[width=\textwidth]{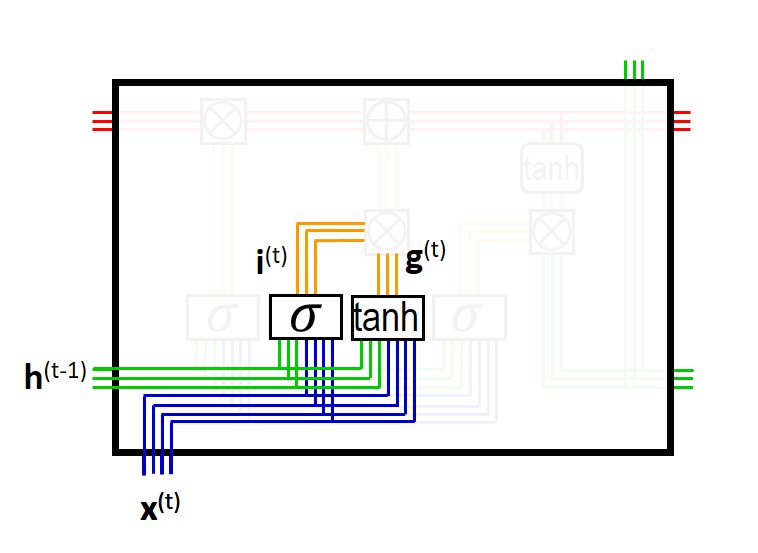}
		\caption{Input gate and input node.}
		\label{figure-LSTM-gates-input-gate}
	\end{subfigure}%
	
	
	\begin{subfigure}[t]{0.5\textwidth}
		\includegraphics[width=\textwidth]{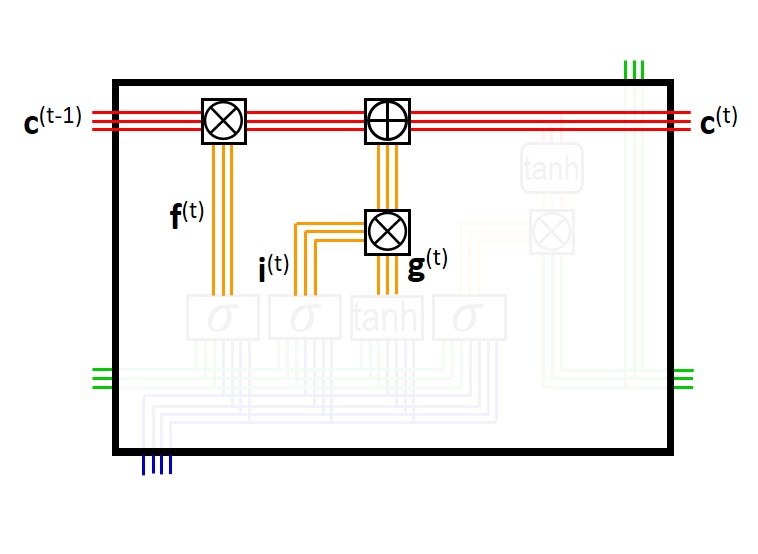}
		\caption{Cell state update.}
		\label{figure-LSTM-gates-cell-state-update}
	\end{subfigure}\hfill%
	\begin{subfigure}[t]{0.5\textwidth}
		\includegraphics[width=\textwidth]{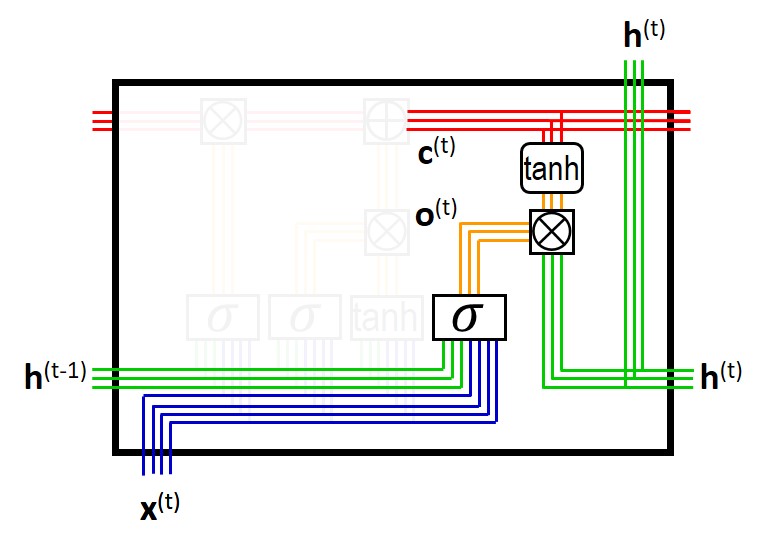}
		\caption{Output gate and hidden state update.}
		\label{figure-LSTM-gates-output-gate}
	\end{subfigure}%
	
	\captionsetup{width=1.0\textwidth}
	\caption{Long short-term memory states and gates.}
	\label{figure-Long-short-term-memory-cell-gates}
\end{figure*}

\noindent
We will now explain how an LSTM works, the main components of the cell (\fref{Figure-LSTM-main-diagram}) and the complete algorithm (Equations \ref{Equation-LSTM-01} to \ref{Equation-LSTM-06}). In each LSTM cell the flow of information is controlled by tree gates, namely: forget, input and output gates. The operations performed at cell level are schematically presented in \fref{figure-Long-short-term-memory-cell-gates}.
All calculations at each gate depend on the current inputs at the same time step and the previous hidden state (at time step t-1). The output from the cell depends on those two variables plus the cell state at time step t. 
%

\subsubsection*{Forget gate}

\noindent
The forget gate defines which information to remove or ``forget" from the cell state. For this purpose, the forget gate has a neural network with a logistic sigmoid activation function ranging from 0 to 1 (\fref{figure-LSTM-gates-forget-gate}). The extremes of that interval correspond to: keep this information (1) or completely remove this information (0). The maths operations performed at this gate are represented by \eref{Equation-LSTM-01}.


\subsubsection*{Input gate and input node}

\noindent
The input gate specifies which information to add to the previous cell state. This part of the cell comprises two elements as shown in \fref{figure-LSTM-gates-input-gate}. The first component is the input gate, where the inputs (hidden state at time step t-1 and inputs at time step t) go through a neural network with a logistic sigmoid activation function (corresponds to node i and function $ \bm{i}^{(t)} $). The second component is the input node (to differentiate from ``gate") and represents the new ``candidates" that could be added to the cell state. These are generated through a neural network with a hyperbolic tangent activation function (corresponds to node g and function $ \bm{g}^{(t)} $). The corresponding operations carried out in this section of the cell are represented by Equations \ref{Equation-LSTM-02} and \ref{Equation-LSTM-03}.


\subsubsection*{Cell state update}

\noindent
The cell state update is performed using the results of both the forget and input gates. The operations implemented for this purpose are presented schematically in \fref{figure-LSTM-gates-cell-state-update} and in mathematical terms by \eref{Equation-LSTM-05}.


\subsubsection*{Output gate and hidden state update}

\noindent
The output gate defines the information from the cell state that will be used as output of the memory cell for the present time step. This gate has the fourth neural network of the LSTM cell, with a logistic sigmoid activation function (\fref{figure-LSTM-gates-output-gate}). The operations applied in this gate are represented by \eref{Equation-LSTM-04}. Finally, the actual output from the cell results from 
the hidden state update, which is computed using \eref{Equation-LSTM-06}, taking into account the results of the output gate and the present cell state, pushed through a hyperbolic tangent function (other activation functions may be used).

\subsubsection*{Cell and hidden states}

\noindent
The information flows through the gates described above, and they are important to justify what happens in the states. 
However, all the information is transmitted to the following time step through the states.
In this sense, the signals from the states summarise all information. 
The main concept behind the cell state is that it represents the long-term memory of the model, while the hidden state corresponds to the short-term memory.

\subsection{Methodology}
\label{section-OpeningTheLSTMBlackBox-Methodology}


\noindent
The analysis of signals in the LSTM states and gates is conducted at different levels of the memory cell, specifically:
forget gate (\eref{Equation-LSTM-01});
product of the outputs from the input gate and input node (Equations \ref{Equation-LSTM-02} and \ref{Equation-LSTM-03});
output gate (\eref{Equation-LSTM-04});
cell state (\eref{Equation-LSTM-05}); and
hidden state (\eref{Equation-LSTM-06}).
The product of the outputs from input gate and input node is chosen instead of the individual outputs. The main reason is that the product is what is added to update the previous cell state, thus having most relevant and interpretable information.
Using the equations referred above, the signals are calculated in each of those locations, 
at every time step,
and for each individual unit of the LSTM memory cell.

Regarding features, three different cases are analysed to examine whether the behaviour found is consistent under different conditions.
In this case we use univariate (feature set 1) and multivariate LSTM models (feature sets 2 and 3). The 10-year bond yield is a feature in all sets considered. 
In the second set we add a technical momentum indicator developed by Merrill Lynch, now Bank of America Merrill Lynch \citep{Garman2001MLHYmodel}. 
This indicator is based on the concept of ``reversion to the mean" and assuming normal distribution of the deviations from a short-term average of 30 working days.
The third feature set includes the 10-year yield and the two closest benchmarks in the yield curve, the 5-year and 30-year yield. 
Yields with maturities adjacent to the one we want to forecast were found to be relevant features in previous work 
\citep{nunes2019eswaPaper}.
A summary of the LSTM model used in this study is presented in \tref{table-Summary-of-empirical-work-signals-analysis}.

\begin{table}[!htb]
	\bigskip
	\centering
	\caption{Summary of the LSTM model used for signal analysis.}
	\label{table-Summary-of-empirical-work-signals-analysis}
	
	\begin{tabularx}{1.0\columnwidth}{l l X}
		
		\toprule
		
		\multicolumn{3}{l}{\textbf{LSTM architecture}} \\
		\midrule
		
		Features                   & set 1       & 10-year yield                              \\
		                           & set 2       & 10-year, momentum indicator                \\
		                           & set 3       & 10-year, 5-year, 30-year yield             \\
		
		\multicolumn{2}{l}{Target}               & 10-year yield                              \\
		\multicolumn{2}{l}{Forecasting horizon}  & next day + 5 days                          \\
		\multicolumn{2}{l}{Moving window size}   & 3000 days                                  \\
		
		\multicolumn{2}{l}{Hidden units}         & 3                                          \\
		\multicolumn{2}{l}{Sequence in}          & 6 days                                     \\
		\multicolumn{2}{l}{Sequence out}         & 6 days                                     \\
		
		\bottomrule
		
	\end{tabularx}
	\bigskip
\end{table}

Additional options for the signals analysis are justified below.
First, for this study we select forecasting horizon of next day + 5 days. The results presented in 
\sref{section-MemoryAdvantageOfLSTMs-Results-and-discussion}
show that for this horizon, there is already a significant differentiation between MLP and LSTM models.
%
Second, the number of hidden unit in the LSTM is reduced to three for interpretability purposes.
%
Third and last, sequence-to-sequence architectures are used with both input and output sequences equal to 6 days.
On the one hand, the input sequence equal to 6 days is in line with the option adopted in 
\sref{section-MemoryAdvantageOfLSTMs-Methodology}.
On the other hand, the output sequence also equal to 6 days insures that the last value of the output sequence corresponds to the forecasting horizon of next day + 5 days we aim to predict.

\subsection{Results and discussion}
\label{section-OpeningTheLSTMBlackBox-Results-and-discussion}

\begin{figure*}[!htbp]
	\centering
	\begin{subfigure}[t]{1.0\textwidth}
		\captionsetup{width=1.0\textwidth}
		\includegraphics[width=\textwidth]{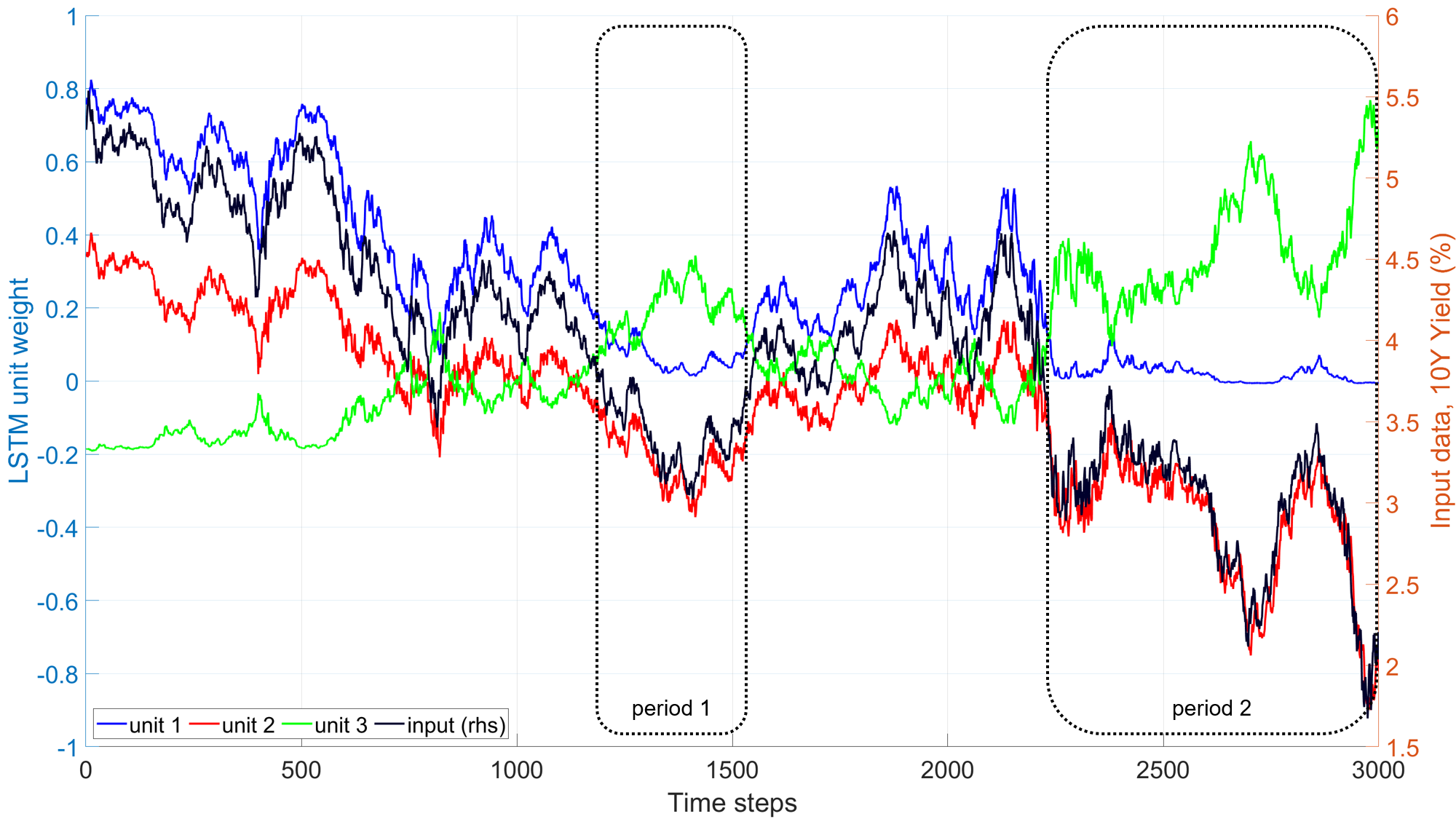}
		\caption{Hidden state signals.}
		\label{figure-Signals_FeatSet1_HiddenState}
	\end{subfigure}
	
	\vspace{0.50cm}
	
	\begin{subfigure}[t]{1.0\textwidth}
		\includegraphics[width=\textwidth]{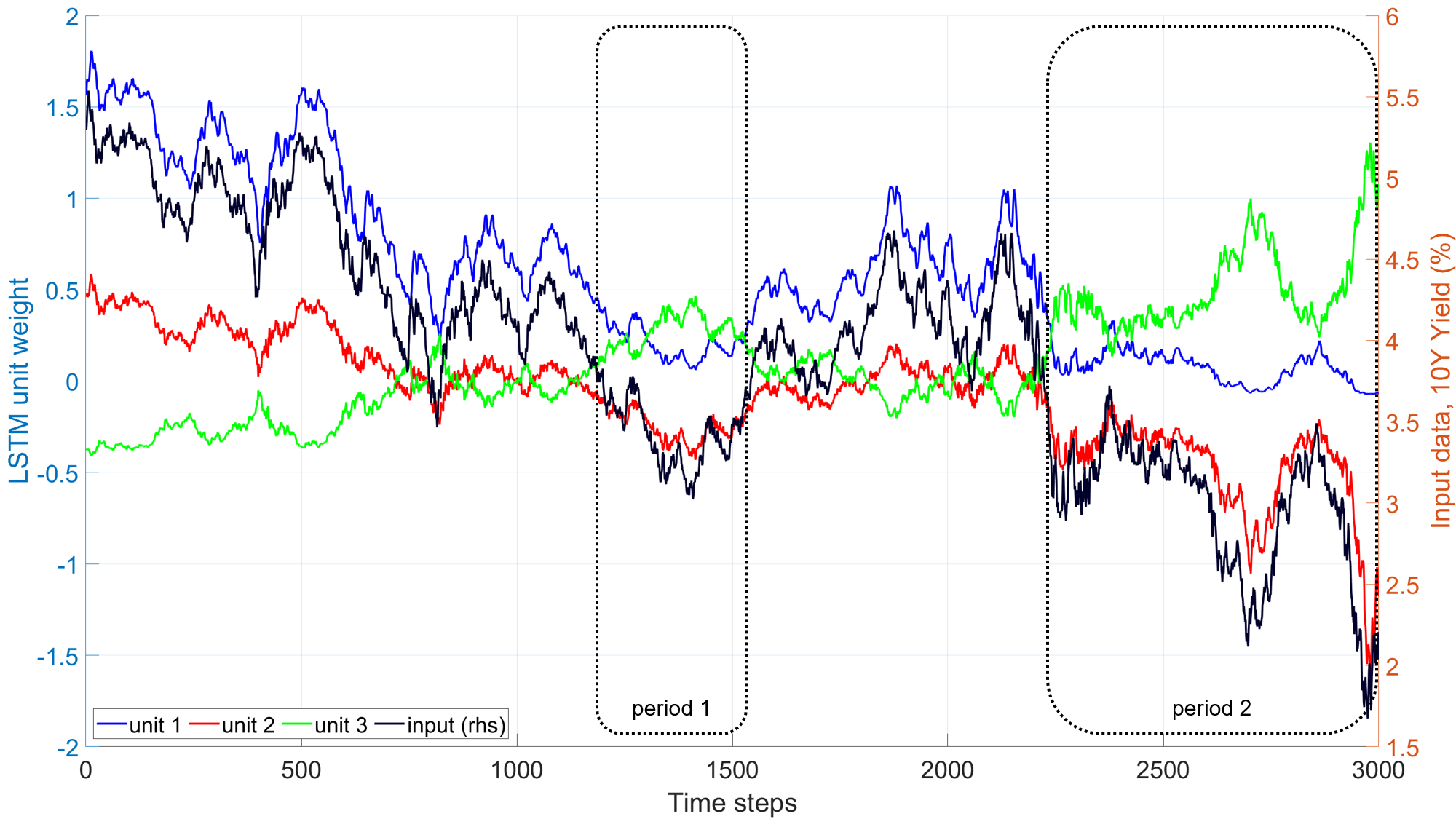}
		\caption{Cell state signals.}
		\label{figure-Signals_FeatSet1_CellState}
	\end{subfigure}
	
	\vspace{0.10cm}
	
	\captionsetup{width=1.0\textwidth}
	\caption{Long short-term memory network signals.}
	\label{figure-Long-short-term-memory-network-signals}
\end{figure*}

\noindent
The main results for the different feature sets are presented in this section, starting with the univariate model (feature set 1).
The signals for both hidden and cell states are presented in 
\fref{figure-Long-short-term-memory-network-signals}.
These signals are plotted against the 10-year yield, for reference and facilitate the interpretation process.

From the results, we can observe some similarity between signals of the hidden and cell states.
It is worth emphasising that the hidden state at time step $ t $ is calculated using the cell state at the same time step and the result of the output gate using \eref{Equation-LSTM-06}.
Consequently, some similarity between those signals may be expected.

A more remarkable property shown in the signals is that unit 1 becomes almost inactive, both in terms of volatility and weight, during two different periods identified as period 1 and 2 in \fref{figure-Long-short-term-memory-network-signals}. 
During those, while unit 1 tends to a zero weight, units 2 and 3 take over becoming more active and following more closely the volatility of the 10-year bond yield.
The periods mentioned are naturally different both in terms of duration and occurrence in time.
However, they both correspond to periods in which the 10-year yield assumes downward extreme values, more specifically yields of 3.6-3.7\% and below.
Taking this into account, there is some evidence that some form of specialisation of the units occur.
In this case, upward values covered by unit 1 and downward extreme values controlled by units 2 and 3.
That is, specialisation of units covering different yield ranges.

\begin{figure*}[!htbp]
	\centering
	\begin{subfigure}[t]{1.0\textwidth}
		\captionsetup{width=1.0\textwidth}
		\includegraphics[width=\textwidth]{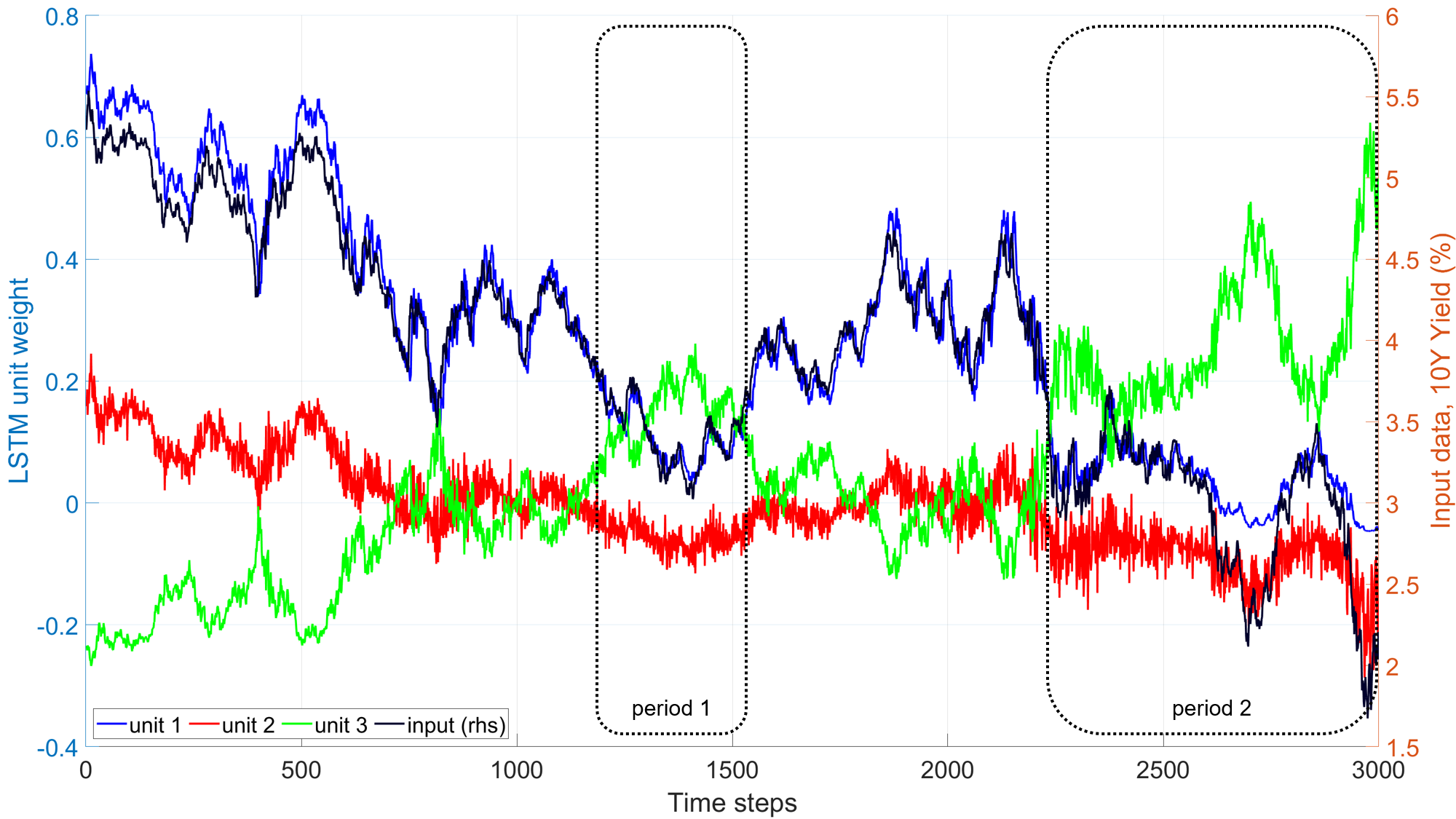}
		\caption{Gate signals: input gate $ \otimes $ input node.}
		\label{figure-Signals_FeatSet1_InputNewCandGate}
	\end{subfigure}
	
	\vspace{0.50cm}
	
	\begin{subfigure}[t]{1.0\textwidth}
		\includegraphics[width=\textwidth]{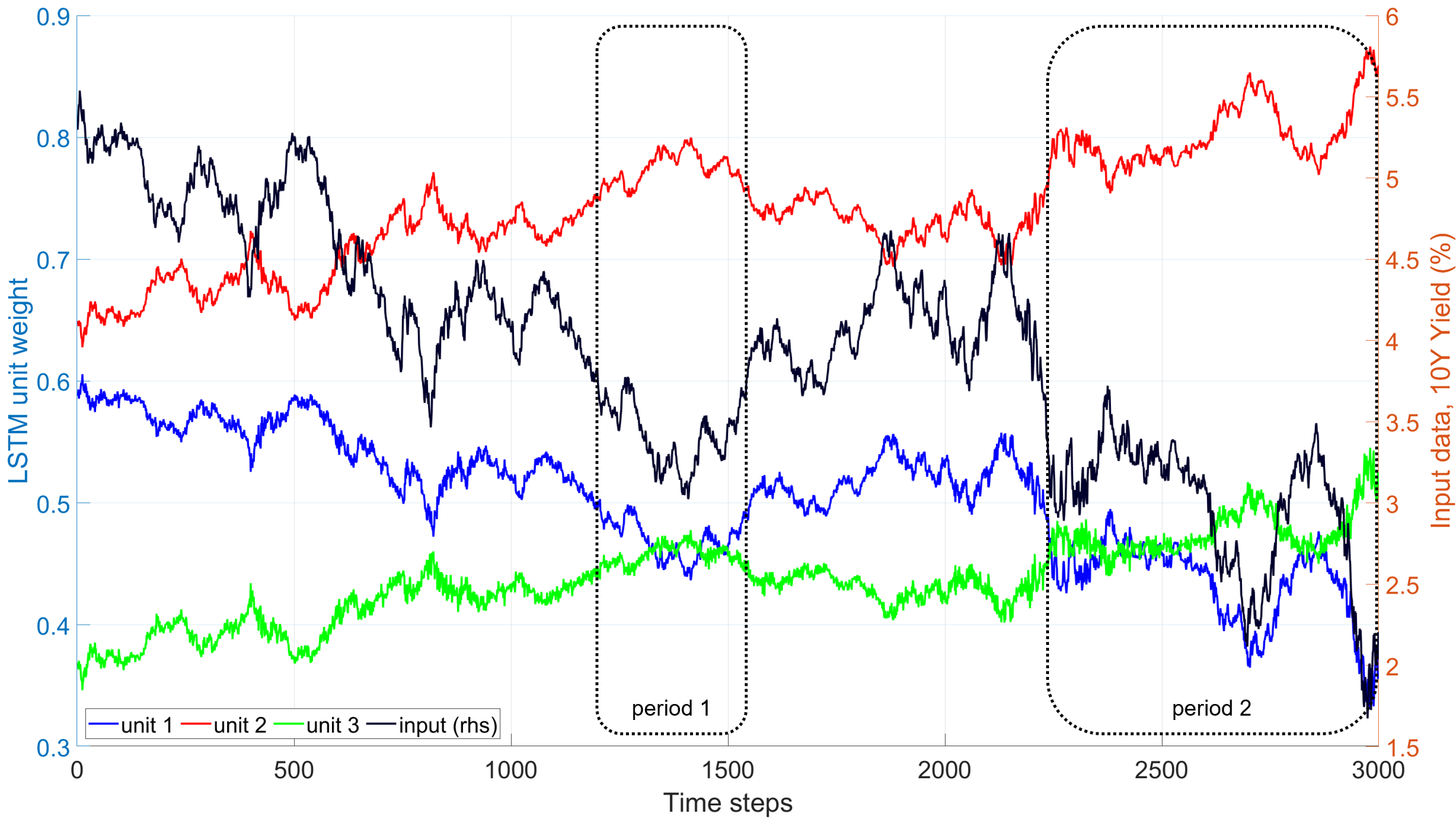}
		\caption{Gate signals: forget gate.}
		\label{figure-Signals_FeatSet1_ForgetGate}
	\end{subfigure}
	
	\vspace{0.10cm}
	
	\captionsetup{width=1.0\textwidth}
	\caption{Example of signals at gates level.}
	\label{figure-Example-of-signals-at-gates-level}
\end{figure*}

As mentioned before, the hidden and cell states are the most important since they summarise all the information going through all the other gates. The signals at the cell gates are helpful to understand and confirm what happens at the states' level.
In this vein, and as an example of this type of confirmation in relation to the cell state, we present the signals at the following two locations in the cell: input gate $ \otimes $ input node and forget gate
(\fref{figure-Example-of-signals-at-gates-level}).
%
%
%
%
We observed that unit 1 becomes almost inactive tending to a zero weight during periods 1 and 2
(\fref{figure-Long-short-term-memory-network-signals}). 
This behaviour is better understood and justified at gate level.
Indeed, at input gate $ \otimes $ input node location the resulting signal is adding approximately zero of unit 1 to the previous cell state
(\fref{figure-Signals_FeatSet1_InputNewCandGate}). 
At the same time, the forget gate is increasing the amount to forget of unit 1 (via lower weights) during the same periods
(\fref{figure-Signals_FeatSet1_ForgetGate}), while decreasing it for the other two units (via higher weights). 

\begin{figure*}[!htb]
	\centering
	\captionsetup{width=1.00\textwidth}
	\includegraphics[width=0.98\textwidth]{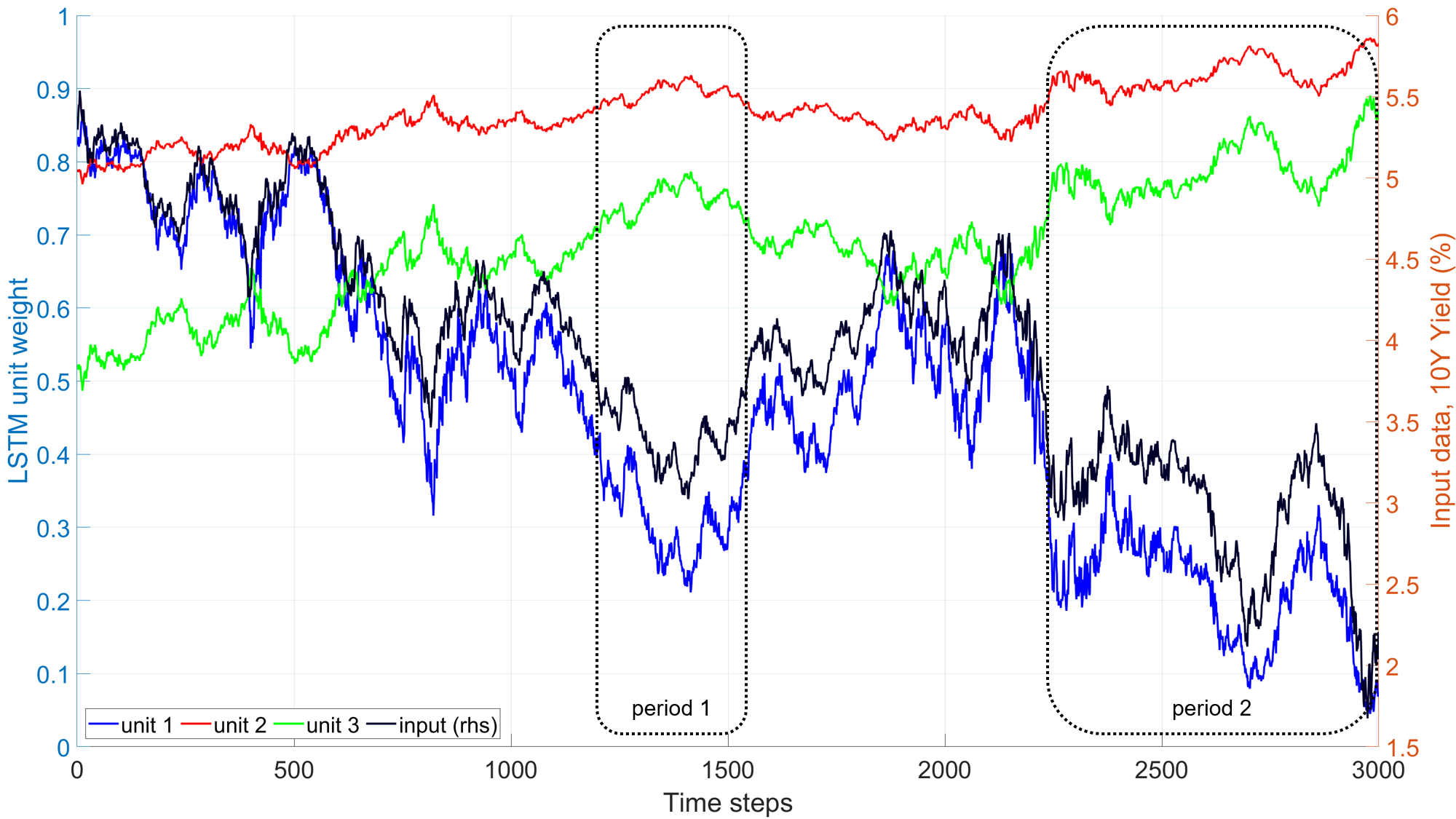}
	\caption{Gate signals: output gate.}
	\label{figure-Signals_FeatSet1_OutputGate}
\end{figure*}

As to the hidden state, the above conclusions are combined with the result from the output gate, which shows a marked decline in the weights for unit 1, with opposite movements for units 2 and 3
(\fref{figure-Signals_FeatSet1_OutputGate}).

%

\begin{figure*}[!htbp]
	\centering
	\begin{subfigure}[t]{1.0\textwidth}
		\captionsetup{width=1.0\textwidth}
		\includegraphics[width=\textwidth]{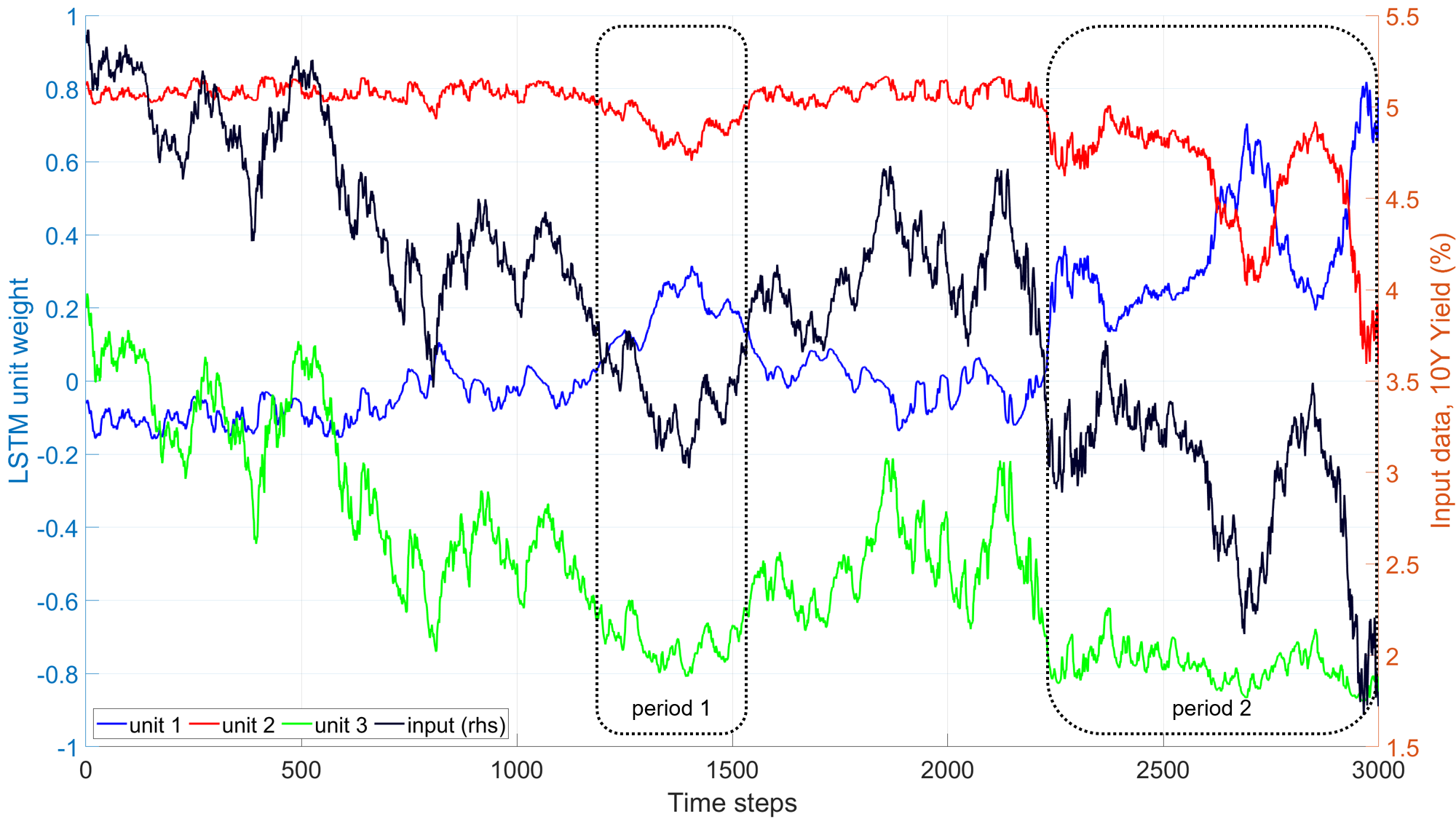}
		\caption{Feature set 2.}
		\label{figure-Signals_FeatSet2_HiddenState}
	\end{subfigure}
	
	\vspace{0.50cm}
	
	\begin{subfigure}[t]{1.0\textwidth}
		\includegraphics[width=\textwidth]{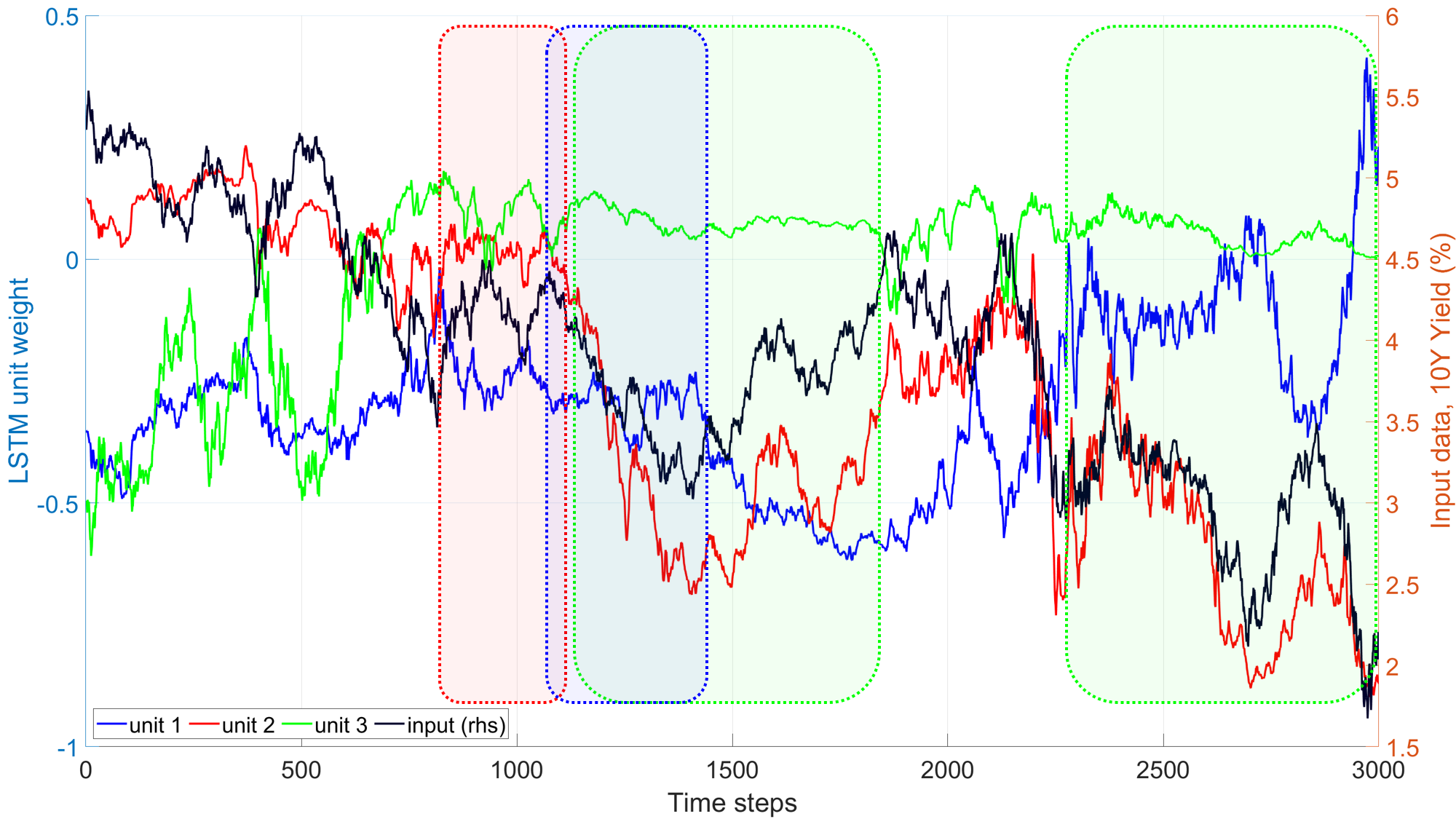}
		\caption{Feature set 3.}
		\label{figure-Signals_FeatSet3_HiddenState}
	\end{subfigure}
	
	\vspace{0.10cm}
	
	\captionsetup{width=1.0\textwidth}
	\caption{Hidden state signals for feature sets 2 and 3.}
	\label{figure-Hidden-states-signals-for-feature-set-2-and-3}
\end{figure*}

Moving to feature sets 2 and 3, although the behaviour is distinct in each case, the same type of activation / deactivation of units can be identified. 
For conciseness, an exemplification of the results obtained for the hidden state for both feature sets is provided in
\fref{figure-Hidden-states-signals-for-feature-set-2-and-3}. 
For set 2 (\fref{figure-Signals_FeatSet2_HiddenState}), despite the high volatility of the second feature 
(momentum indicator, \tref{table-Summary-of-empirical-work-signals-analysis}),
the same two periods can clearly be observed as described for feature set 1
(\fref{figure-Signals_FeatSet1_HiddenState}). 
To note, in this case it is unit 3 that assumes the role of previous unit 1. This switch has no relevance since they are all equal units at the start of the learning process.
Another interesting observation (not presented here for succinctness), is the fact that given the high volatility of the momentum indicator, two of the units tend to ``specialise" in this feature and only one unit in the 10-year yield.

For feature set 3 (\fref{figure-Signals_FeatSet3_HiddenState}), 
the pattern of the signals in relation to the 10-year yield data is much looser. 
One of the reasons that may justify this behaviour is the higher correlation among all yields considered as features (5, 10 and 30-year yield).
As a result, there is a lower level of dependency on only one of them.
Similarly to what was found in the previous feature sets, we can identify some form of yield range specialisation of the units. 
In this case, unit 3 seems to cover a yield range above 4.5\% approximately, becoming much less active subsequently 
(\fref{figure-Signals_FeatSet3_HiddenState}).

Overall, the most remarkable property found consistently in the LSTM signals, for all feature sets,
is the activation / deactivation of units during learning.
It can be characterised by 
an alternation of periods of weight close to zero or low variability, with no significant change in the states,
with periods where the units become highly active giving higher contributions to the forecasting process.
Furthermore, we found evidence that the LSTM units may specialise in different yield ranges or features considered.

\section{New LSTM-LagLasso method}
\label{section-New-LSTM-LagLasso-method}


\noindent
Our intention here is to interpret the representations learned by the LSTM model, whose estimation is driven by purely statistical considerations of the error being minimized, in terms of external effects that might have an influence on the bond yields. Should the model extract meaningful features during the learning process, the representations would correlate with exogenous information that was not available to the learning algorithm. In other words, it is precisely because the representations may match such exogenous information that the predictive ability of the black-box models is good. 

%

\subsection{Methodology}
\label{section-NewLSTMLagLassoMethod-Methodology}

\noindent
In this section, we briefly present the Lasso, the Kalman-LagLasso and then evolve to introduce our methodology, the LSTM-LagLasso.

\subsubsection*{Lasso}

\noindent
The Lasso regression we formulate to explain the signals within the LSTM (features extracted by the LSTM) in terms of exogenous variables has the form \citep{tibshirani1996regression}: 

\begin{equation}
	\label{eq-LassoRegression}
	\underset{w}{min} ~ \{ ~ \| \bm{X} ~ \bm{w} - \bm{s}_{lstm} \|_2^2 + \gamma ~ \| \bm{w} \|_1 ~ \}
\end{equation}

\noindent
where 
$ \bm{X} $			is the matrix of features and respective lags;
$ \bm{w} $      	the vector of unknown parameters;
$ \bm{s}_{lstm} $	are the LSTM cell and hidden state signals (target vectors);
$ \gamma $ 			is the regularisation parameter;
$ \| ~ \|_1 $ 		denotes the $ L_1 $-norm; and
$ \| ~ \|_2 $ 		the $ L_2 $-norm.

The Lasso regression determines the parameters of the model by minimising the sum of squared residuals, using an $ L_1 $-norm penalty for the weights. Due to the type of constraint, it tends to lead to sparse solutions, i.e. some coefficients are exactly zero and as a result the corresponding features are discarded. This is particularly important since it enables a continuous type of feature selection through the tuning of the regularisation parameter $\gamma$, and the identification of the most relevant features for the model.

\subsubsection*{Kalman-LagLasso}

\noindent
\citet{mahler2009modeling} introduced the Kalman-LagLasso method in a study to predict the monthly changes of the S\&P500 index, using macroeconomic and financial variables. 
The overall procedure included two phases. In the first, Kalman filters \citep{kalman1960KalmanFilterOriginal, niranjan1996sequential} were used to denoise the explanatory variables and predict the residuals (part not explained by the model). 
Then the LagLasso phase is implemented, to determine the most relevant features and respective lag using the filtered/denoised variables to explain the prediction residuals.
The LagLasso method is based on Lasso \citep{tibshirani1996regression}, implemented via the modified Least Angle Regression (LARS) algorithm \citep{efron2004LARS}.
The algoritm was modified by \citet{mahler2009modeling}, so that only one lag per feature is selected. 
Once selected, the other lags of the same feature are eliminated from the active set of features so that they cannot be selected again.
The Kalman-LagLasso is an elegant method of combining a forecasting method with error analysis, seeking to explain the part not explained by model, i.e. the residuals, with external financial variables.

More recently \citep{montesdeoca2019bitcoin}, and using a similar approach, the Kalman-LagLasso method has been used to compare the type of information influencing US stock indices (S\&P 500 and Dow Jones Industrial Average) in contrast with that affecting cryptocurrencies (Bitcoin and Ethereum).

\subsubsection*{LSTM-LagLasso}

\noindent
The LSTM-LagLasso method is the new methodology we developed 
to explain the signals extracted from the LSTM states.
It is inspired in the Kalman-LagLasso method, 
but with significant modifications in terms of model used, 
target variable to which the LagLasso method is applied, 
as well as the methodology to determine the relevant features and respective lags.

When compared to the Kalman-LagLasso method,
our objective is different and we aim to analyse what information is contained in the signals, in particular, if they can be explained by external variables.
As a result of this objective, instead of using a Kalman filter implementation of a linear autoregressive model, we use as the main model non-linear LSTMs.

Our methodology also diverges in the target used for the LagLasso procedure. While in the Kalman-LagLasso the target time series is the Kalman residuals, in our methodology we use the signals extracted from the LSTM states. We model the cell and hidden states independently of each other.

On the LagLasso technique, we also differ from the original paper \citep{mahler2009modeling} in several aspects.
%
First, we do not denoise the features and use Lasso directly with the variables and respective lags.
To explain this option, we need to mention that
Kalman filters are often applied to sensors data \citep{park2019KalmanFilterSensors}, 
and it is known that this type of data does not provide exact readings/measurements.
This is because they introduce their own distortions and
they are always corrupted by noise \citep{maybeck1979KalmanFilterBook}.
In this context, the use of Kalman filter to remove the noise of sensor data is a natural application.

However, in financial markets the concept of noise cannot realistically be applied, in our opinion.
Instead, it is the complexity of market dynamics that is responsible for price formation.
Indeed, it is the multiplicity of factors influencing the markets that makes the problem extremely complex.
Part of those factors are already incorporated in the historic values of the time series, 
but another important part is coming from new information arriving to the markets in real-time, 
together with the new expectation and reactions of market participants to that new information.
Ultimately, from the interaction of all these factors results a new equilibrium and a new real-time asset valuation.
This concept applied more directly to market variables, can also be extended to macroeconomic indicators. 
Our approach seems to be more appropriate when dealing with financial time series.

The second main difference in our methodology is that we consider all lags selected as relevant in the LSTM-LagLasso algorithm and not only one for each feature, as proposed in the Kalman-LagLasso procedure \citep{mahler2009modeling}. We found no reason to limit the number of lags that may participate in the forecasting process to only one.
The LSTM-LagLasso methodology is outlined in \alref{algo-LagLassoAlgorithm}.

\IncMargin{0.5em}
\begin{algorithm}[htbp]
	\caption{LSTM-LagLasso algorithm}
	\label{algo-LagLassoAlgorithm}

	\SetAlgoLined
	\SetAlgoLongEnd              
	\DontPrintSemicolon
	
	\SetKwData{Left}{left}\SetKwData{This}{this}\SetKwData{Up}{up}
	\SetKwFunction{Union}{Union}\SetKwFunction{FindCompress}{FindCompress}
	\SetKwInOut{Input}{input}\SetKwInOut{Output}{target}
	\Input{Macroeconomic and market features}
	\Output{LSTM cell and hidden state signals}
	\BlankLine

	Extract LSTM cell and hidden state signals and set them as targets in independent models (one per state).\;
	Select the number of lags to consider $\{k \in [1,6]\}$.
	\label{algo-LagLassoAlgorithm-NumLags}\;
	Build matrix $\bm{M}$ containing macroeconomic and market features.
	\label{algo-LagLassoAlgorithm-FeaturesMatrix}\;
	Transform matrix $\bm{M}$ into matrix $\bm{X}$ by incorporating $k$ lags.\;
	Standardize all variables (features and target) to have zero mean and unit standard deviation.\;
	
	\For{$i\leftarrow \gamma_{min}$ \KwTo $\gamma_{max}$}{
		Perform Lasso regression (\eref{eq-LassoRegression}).\;
		Identify non-zero weight values in vector $\bm{w}$.\;
	}
	
	Select $\gamma$ (look for stabilising trend in the number of features against $\gamma$ and forecasting error).
	\label{algo-LagLassoAlgorithm-selectGamma}\;
	Perform Lasso regression with selected $\gamma$.\;
	Identify the final most relevant variables and lags (non-zero weight values in vector $\bm{w}$).\;

\end{algorithm}
\DecMargin{0.5em}

\begin{figure}[!htb]
	\centering
	\includegraphics[width=0.475\textwidth]{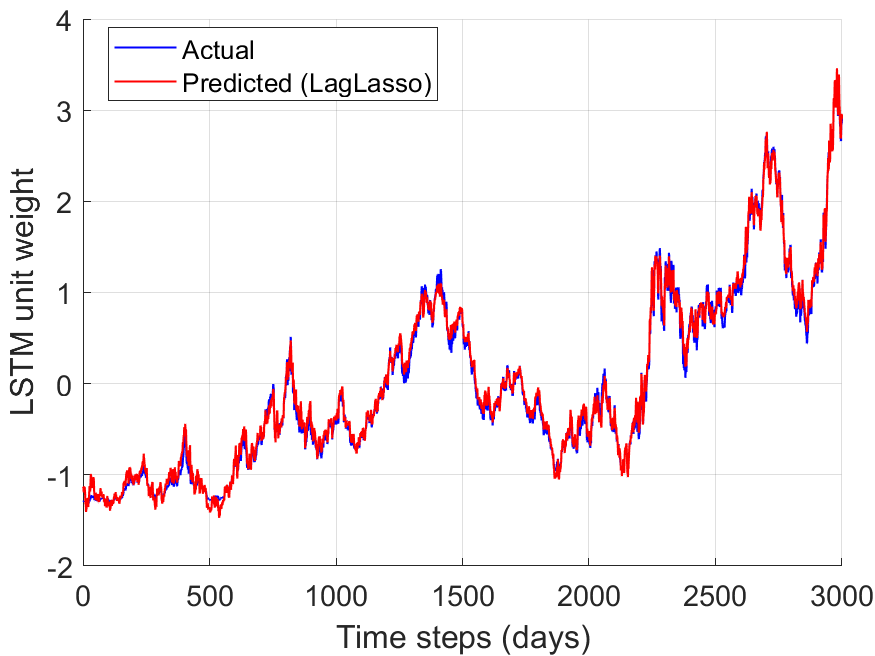}
	
	\captionsetup{width=0.475\textwidth}
	\caption{Example of LagLasso prediction of LSTM signals: actual versus predicted, for the hidden state, unit 3.}
	\label{figure-LagLasso_Actual_Predicted_excl10Y_State1_Unit3}
\end{figure}

The LSTM-LagLasso method is applied individually to both hidden and cell states, and for each of the three LSTM units.
Additional clarification of the options considered are presented below. 
First, the number of lags is equal to those of the sequences (input and output) used in the LSTM model in
\sref{section-OpeningTheLSTMBlackBox-Methodology}, i.e., six lags per feature
(\alref{algo-LagLassoAlgorithm}, Line \ref{algo-LagLassoAlgorithm-NumLags}).
Second, for the external variables to explain the hidden and cell state signals,
we use the same large set of features considered for the MLP model
(\sref{section-MemoryAdvantageOfLSTMs-Data}), 
a list of 159 macroeconomic and market variables
not available to the model during the learning process
(\alref{algo-LagLassoAlgorithm}, Line \ref{algo-LagLassoAlgorithm-FeaturesMatrix}).
Third, for selection of the regularisation parameter we considered both the trend in the number of features against $\gamma$ and the forecasting error
(\alref{algo-LagLassoAlgorithm}, Line \ref{algo-LagLassoAlgorithm-selectGamma}). 
After an initial rapid drop in the number of features, the trend changes significantly, but a clear period of stabilisation cannot be observed. Additionally, for $\gamma$ above 1.0, the error starts increasing more rapidly and the quality of fit deteriorating. For this reason we selected $\gamma = 1.0$ for the final Lasso run, as a good compromise between stabilisation and quality of fit. \fref{figure-LagLasso_Actual_Predicted_excl10Y_State1_Unit3} shows an example of LagLasso prediction of LSTM signals with this option for $\gamma$.
Forth and last, we exclude from the original features the 10-year bond yield, since this information is known to the model, both last available value and respective lags, as the only feature considered in the univariate LSTM model used in \sref{section-LSTM-networks-for-bond-yield-forecasting}.

\subsection{Results and discussion}
\label{section-NewLSTMLagLassoMethod-Results-and-discussion}

\begin{sidewaysfigure*}[!htbp]
	\centering
	\captionsetup{width=0.70\textwidth}
	\includegraphics[height=0.55\textheight]{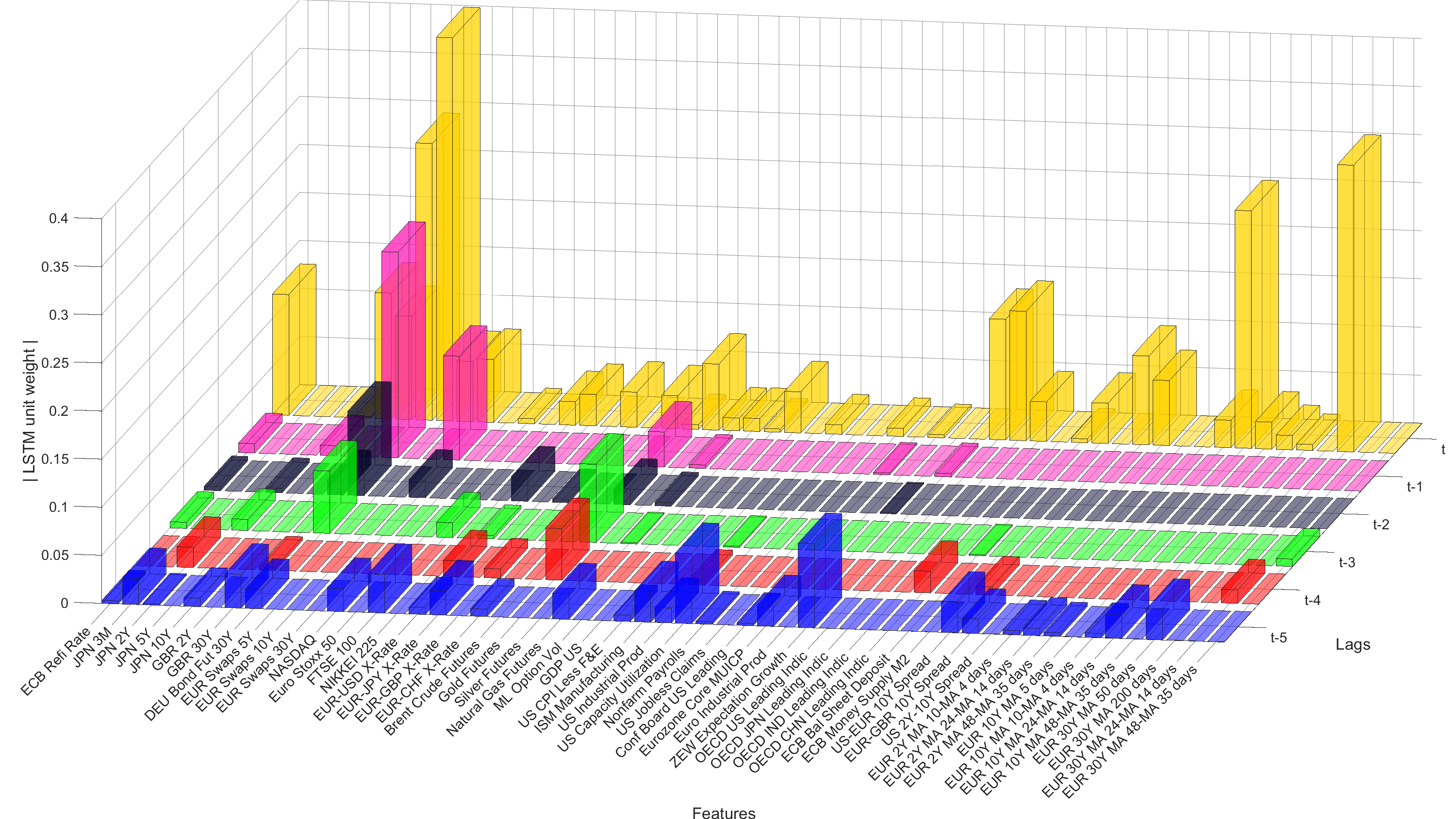}
	\caption{LSTM-LagLasso relevant features for the hidden state, unit 1, considering a regularisation paramete $\gamma$ equal to 1.0.}
	\label{Figure-LagLasso_RelevFeatures_excl10Y_State1_Unit1_Gama1}
\end{sidewaysfigure*}

\begin{sidewaysfigure*}[!htbp]
	\centering
	\captionsetup{width=0.70\textwidth}
	\includegraphics[height=0.55\textheight]{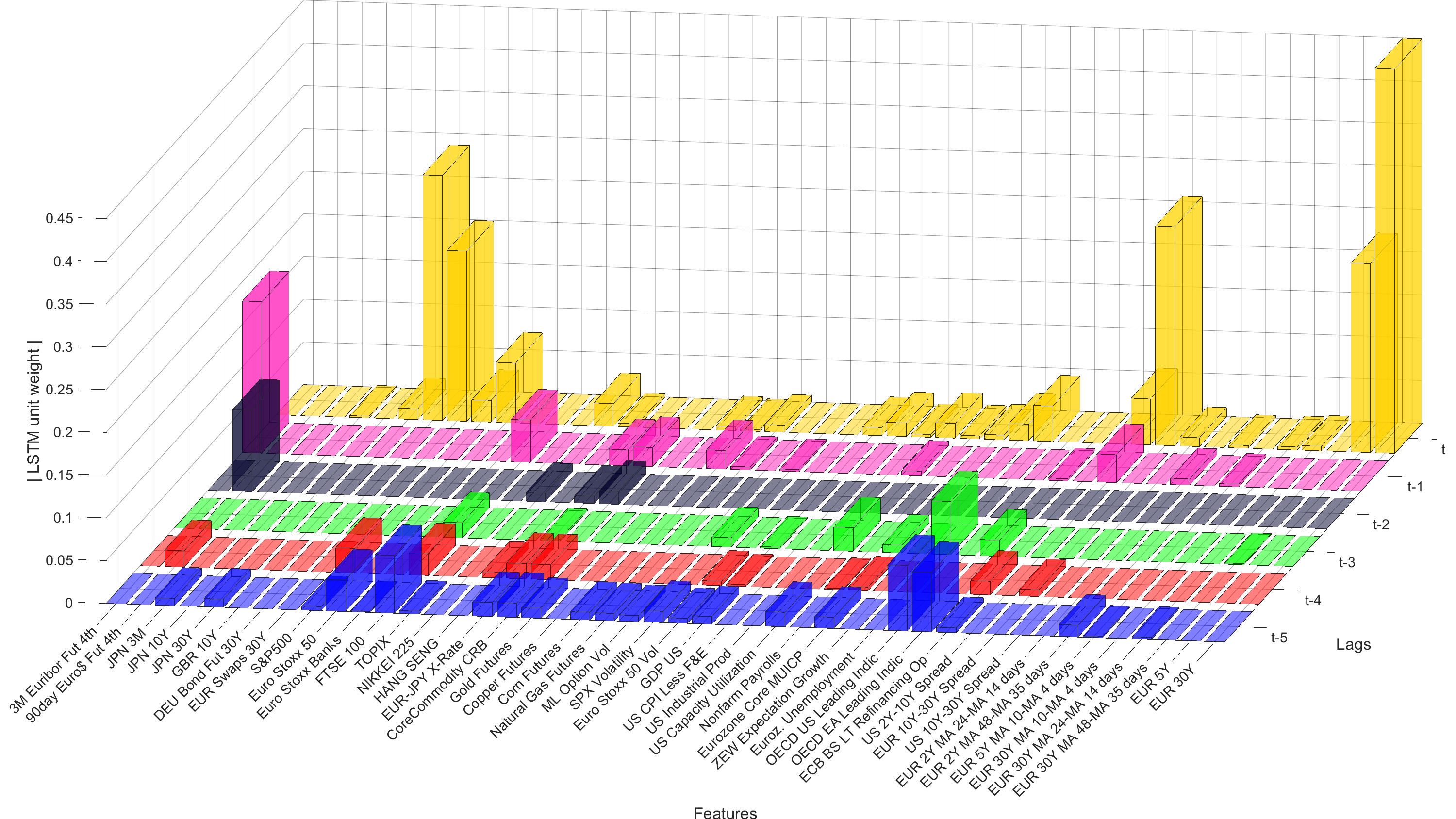}
	\caption{LSTM-LagLasso relevant features for the hidden state, unit 2, considering a regularisation paramete $\gamma$ equal to 1.0.}
	\label{Figure-LagLasso_RelevFeatures_excl10Y_State1_Unit2_Gama1}
\end{sidewaysfigure*}

\begin{sidewaysfigure*}[!htbp]
	\centering
	\captionsetup{width=0.70\textwidth}
	\includegraphics[height=0.55\textheight]{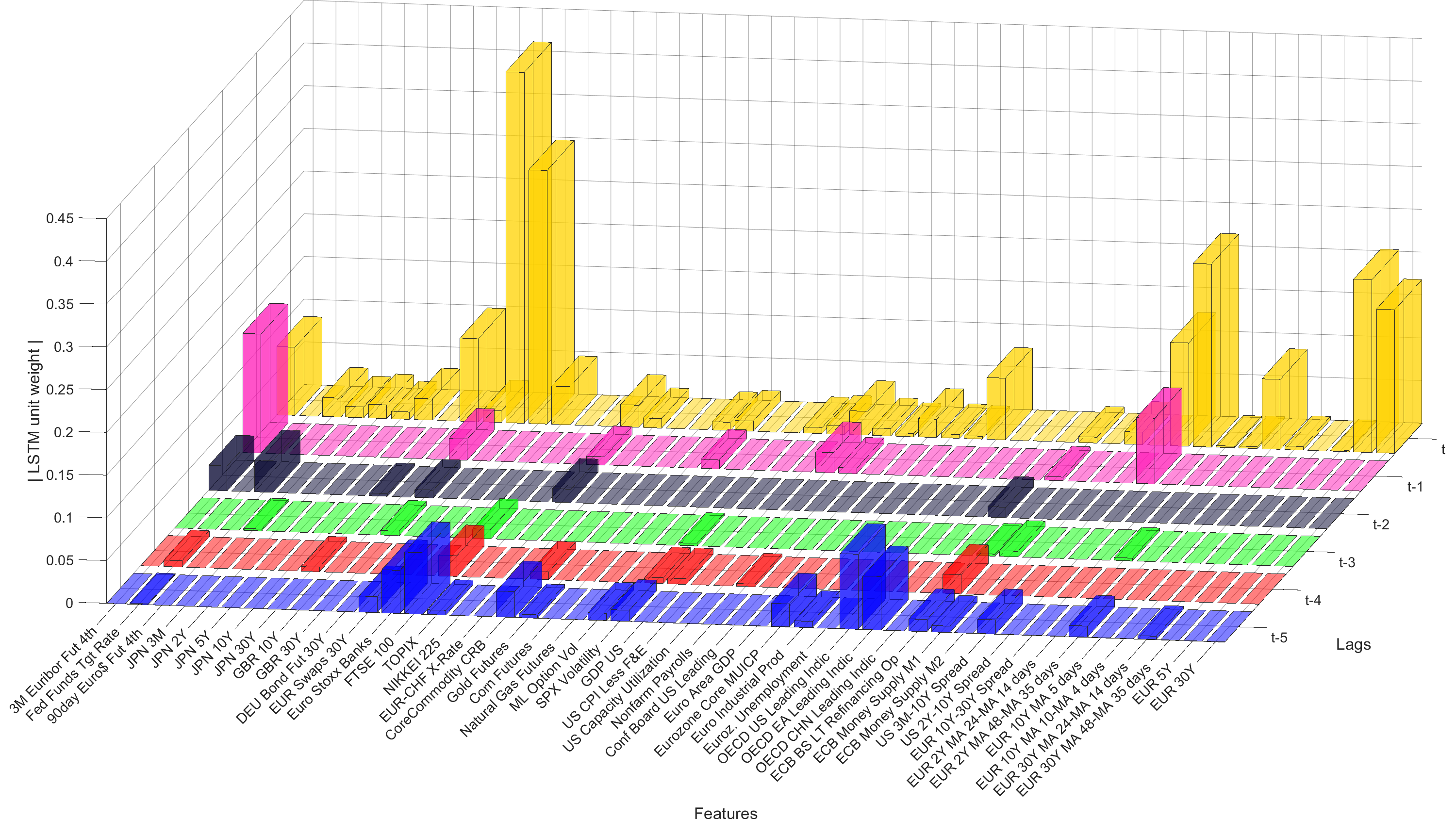}
	\caption{LSTM-LagLasso relevant features for the hidden state, unit 3, considering a regularisation paramete $\gamma$ equal to 1.0.}
	\label{Figure-LagLasso_RelevFeatures_excl10Y_State1_Unit3_Gama1}
\end{sidewaysfigure*}

\noindent
The application of LSTM-LagLasso to the hidden state is presented in Figures \ref{Figure-LagLasso_RelevFeatures_excl10Y_State1_Unit1_Gama1}--\ref{Figure-LagLasso_RelevFeatures_excl10Y_State1_Unit3_Gama1}, 
respectively for hidden units 1 to 3.
The figures show the most relevant variables identified by this methodology.
Note that we plot the absolute value of the weights and not the weights directly, for easier visualisation of the figure.
Since we are using market variables to explain the LSTM signals and not a financial asset target directly, the visualisation of whether the weights are positive of negative is not as relevant as the magnitude of it, i.e. the relevance of the feature as explanatory variable.

\begin{figure}[!htb]
	\centering
	\includegraphics[width=0.475\textwidth]{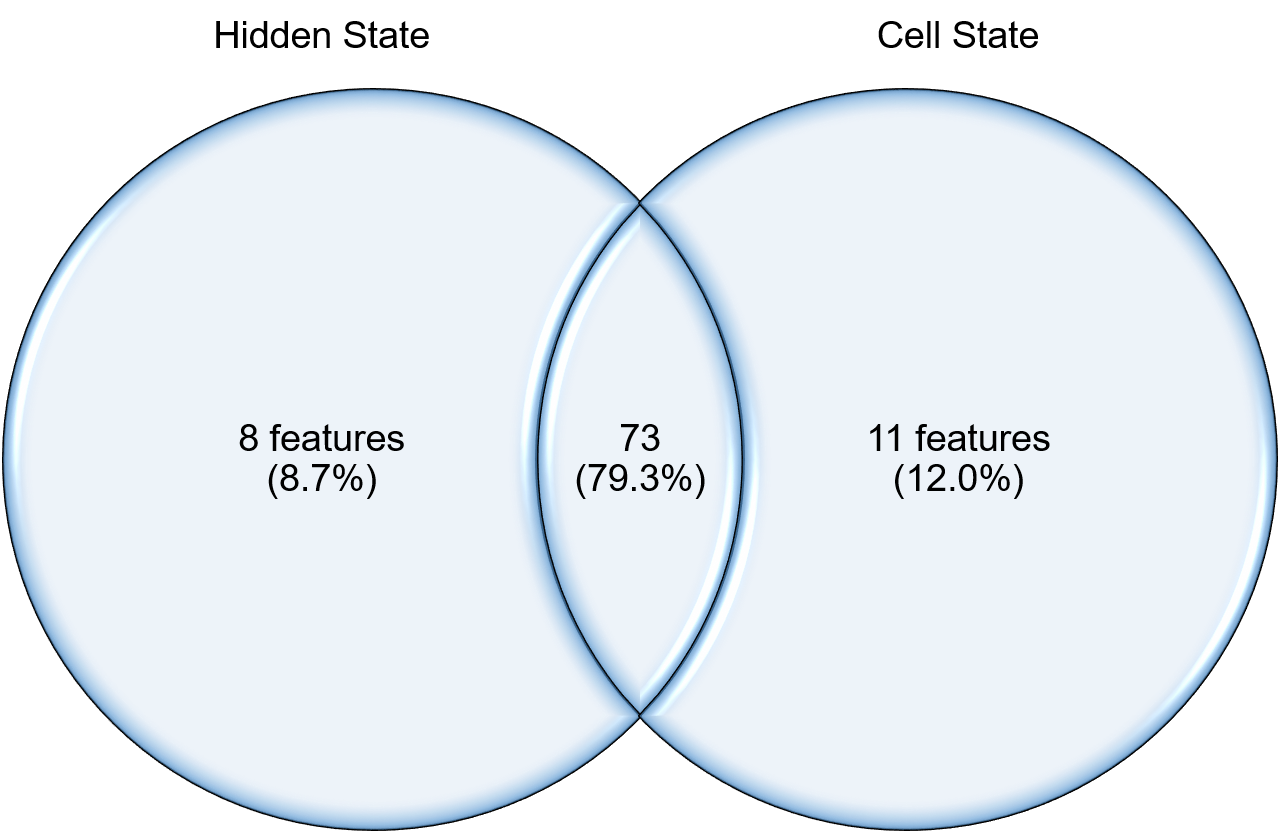}
	
	\captionsetup{width=0.475\textwidth}
	\caption{Comparative Venn diagram of the most relevant features for both hidden and cell states.}
	\label{figure-LagLasso_VennDiagram_excl10Y_State1-2_UnitsAll_Gama1}
\end{figure}

The corresponding results regarding the application of LSTM-LagLasso to the cell state are not presented here, for brevity, but also because the main points that can be extracted from them only reinforce the conclusions for the hidden state.
This can also be observed in the Venn diagram shown in 
\fref{figure-LagLasso_VennDiagram_excl10Y_State1-2_UnitsAll_Gama1},
where almost 80\% of the relevant features identified are common to both hidden and cell state. 
An additional point worth noticing is that the cell state needs slightly more explanatory variables (three more).

\begin{figure}[!htb]
	\centering
	\includegraphics[width=0.475\textwidth]{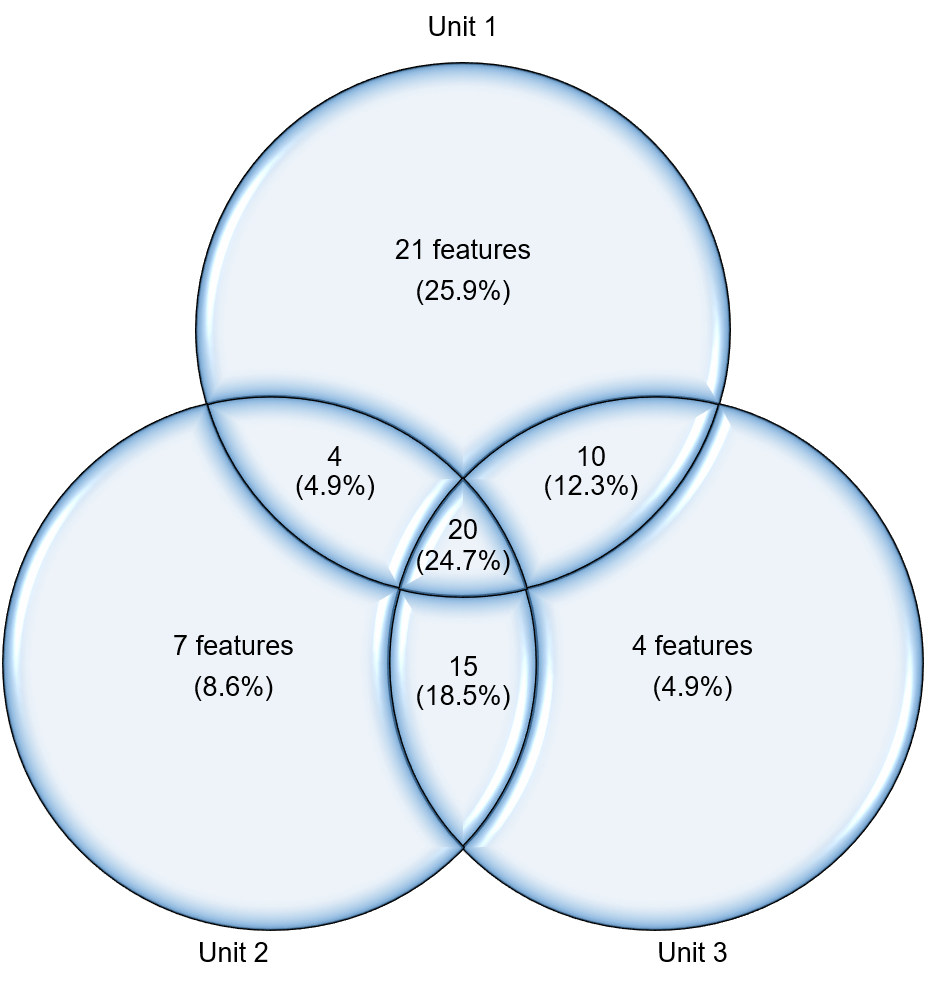}
	
	\captionsetup{width=0.475\textwidth}
	\caption{Venn diagram of the most relevant features for the hidden state and all LSTM hidden units.}
	\label{figure-LagLasso_VennDiagram_excl10Y_State1_UnitsAll_Gama1}
\end{figure}

A similar type of comparison is conducted among the different hidden units, to illustrate the common relevant features identified. The diagram is presented in \fref{figure-LagLasso_VennDiagram_excl10Y_State1_UnitsAll_Gama1}. 
Here we can see that not all top relevant features are common to all hidden units, although 24.7\% of them are common to all three units in the hidden state and 21.4\% in the cell state.

An additional summary of results for the most influential relevant variables, is presented in 
\tref{table-Summary-of-most-influential-relevant-features-weight-higher-than-015}.
For this purpose, only those with an absolute weight greater than or equal to 0.15 are selected.

\begin{table}[!htb]
	\bigskip
	\centering
	\caption{Summary of most influential relevant features for both states. The listed features have an absolute weight greater than or equal to 0.15 in at least one of the LSTM units. The cells in the table for which the weight is equal to zero are left empty.}
	\label{table-Summary-of-most-influential-relevant-features-weight-higher-than-015}
	
	\begin{tabularx}{1.0\columnwidth}{X c c c}
		
		\toprule
		\textbf{Feature name}	& \multicolumn{3}{c}{\textbf{Weight}}						\\
		\cmidrule{2-4}
								& \textbf{Unit 1}	& \textbf{Unit 2}	& \textbf{Unit 3}	\\
		\midrule
		\multicolumn{4}{l}{\textbf{Hidden state}}											\\
		\midrule

		ECB Refi Rate 			& 0.150 			&  					&  					\\
		3M Euribor Fut 4th 		&  					& 0.177 			& 0.248 			\\
		GBR 30Y 				& 0.161 			&  					& 0.053 			\\
		DEU Bond Fut 30Y 		& 0.674 			& 0.287 			& 0.410				\\
		EUR Swaps 5Y 			& 0.398 			&  					&  					\\
		EUR Swaps 30Y 			& 0.192 			& 0.199 			& 0.315 			\\
		FTSE 100 				& 0.056 			& 0.160				& 0.107 			\\
		Gold Futures 			& 0.193 			& 0.012 			& 0.004 			\\
		US 2Y-10Y Spread 		& 0.068 			& 0.087 			& 0.198 			\\
		EUR 10Y-30Y Spread 		&  					& 0.264 			& 0.214 			\\
		EUR 10Y MA 5 days 		& 0.247 			&  					& 0.082 			\\
		EUR 30Y MA 200 days 	& 0.298 			&  					&  					\\
		EUR 5Y 					&  					& 0.221 			& 0.202 			\\
		EUR 30Y 				&  					& 0.450				& 0.168				\\

		\midrule
		\multicolumn{4}{l}{\textbf{Cell state}}												\\
		\midrule

		3M Euribor Fut 4th 		&  					&  					& 0.212 			\\
		90day Euro\$ Fut 5th 	&  					& 0.203 			&  					\\
		GBR 30Y 				& 0.151 			& 0.038 			&  					\\
		DEU Bond Fut 30Y 		& 0.723 			& 0.921 			& 0.533 			\\
		EUR Swaps 5Y 			& 0.258 			&  					&  					\\
		EUR Swaps 30Y 			& 0.220				&  					& 0.186 			\\
		Gold Futures 			& 0.168 			& 0.109 			& 0.051 			\\
		EUR 10Y-30Y Spread 		&  					& 0.271 			& 0.270				\\
		EUR 10Y MA 5 days 		& 0.525 			&  					&  					\\
		EUR 30Y MA 200 days 	& 0.273 			&  					&  					\\
		EUR 5Y 					&  					& 0.003 			& 0.172 			\\
		EUR 30Y 				&  					&  					& 0.240				\\
		
		\bottomrule
		
	\end{tabularx}
	\bigskip
\end{table}

From the results a number of conclusions can be drawn
(Figures \ref{Figure-LagLasso_RelevFeatures_excl10Y_State1_Unit1_Gama1}--\ref{Figure-LagLasso_RelevFeatures_excl10Y_State1_Unit3_Gama1}
and \tref{table-Summary-of-most-influential-relevant-features-weight-higher-than-015}).
%
First, the results confirm that the signals can be explained by external sources of information, not available to the models both during training or forecasting.
%
Second, the LSTM signals are complex and require a significant number of explanatory variables.
%
Third, we observe that for many features there are several lags that are relevant for the prediction process. 
The most important lags are the last two values (t, t-1) and the lag one week before (t-5). 
The importance of the latter lag is interesting, pointing to some type of weekly seasonality or influence.
Given this conclusion that lags are important, selecting only one lag per feature, as proposed in the Kalman-LagLasso method \citep{mahler2009modeling}, would eliminate this additional information, limiting the forecasting ability.

Fourth and last, using the LSTM-laglasso some of the most relevant features selected are conventional market/macro variables, but others are less common, non-conventional ones. We refer here ``conventional" in the sense that they have been used more frequently in modelling financial assets in the past
\citep{nelson1987parsimonious, dunis2007economic, arrieta2015testing}, 
or are more common sense variables for that purpose.

In the conventional group of relevant features, we can refer those related to
central bank reference rates								
(ECB refinancing rate), 
macroeconomic indicators of inflation						
(US Consumer Price Index less food and energy, 
Eurozone Core Monetary Union Index of Consumer Prices),
economic growth / growth expectations						
(Institute for Supply Management Manufacturing,
US Industrial Production,
US Capacity Utilisation,
ZEW Eurozone Expectation of Economic Growth),
and labour market											
(Eurozone Unemployment).

But the explanatory variables go well beyond that group, with a wide range of relevant features in the non-conventional group.
Some of them are specific to the bond market, all of them adding significant information to the previous group.
%
The top relevant feature by weight is the long German government bond future (DEU Bond Fut 30Y),
reaching the value of 0.921 for unit 2 of the cell state
(\tref{table-Summary-of-most-influential-relevant-features-weight-higher-than-015}). 
Note that contrary to what happens at the gates (output of a sigmoid function between 0 and 1), 
the weights in the LSTM states are not limited to 1.
%
Also within the futures asset class, we find the 
3M Euribor and 90d Euro\$ Futures (4th and 5th contracts). 
These contracts have a horizon of approximately one year ahead, 
thus incorporating investors' expectations on the evolution of short-term rates.		

In addition, financial instruments with maturities adjacent to the one we want to predict are also included in this group of relevant features, namely: 
5 and 30-year Euro government bond yield (note that we have excluded the 10-year yield from the LSTM-LagLasso set of features as mentioned in \sref{section-NewLSTMLagLassoMethod-Methodology}); 
2, 10 and 30-year UK government bond yield; and 
5, 10 and 30-year EUR swap rates.
%
Directly related with yields, we can identify as relevant features 
intra-curve spreads such as
US 2--10-year spread and EUR 10--30-year spread; as well as
inter-curve spreads, specifically, 
EUR-GBR 10-year spread and EUR-JPN 10-year spread.

Furthermore, the most relevant features determined via LSTM-LagLasso also include the following asset classes and macroeconomic variables:
commodities (Gold Futures and Brent Crude Futures); 
equity indices (Euro Stoxx 50, FTSE100, and S\&P500);
foreign exchange rates (EUR-USD X-Rate and EUR-JPY X-Rate);
the ECB Balance Sheet Long-Term Refinancing Operations;
OECD Leading Indicators of US, European Area, and Japan; and finally
technical analysis indicators (5-year, 10-year, 30-year moving averages of 5, 50 and 200 days).

It is important to emphasise some aspects regarding the latter group of non-conventional relevant features identified using the LSTM-LagLasso.
%
First, the 5-year and 30-year are adjacent maturities to the 10-year yield we are studying 
and tend to lead flattening and steepening movements of the yield curve around the 10-year maturity.
In particular the long German government bond future is a leveraged instrument with very long maturity and duration. Consequently, they are highly price-sensitive and react very quickly to market movements. This justifies being a top relevant feature.
The second aspect worth highlighting is that, contrary to what could be expected, most of the non-conventional features have higher weights than the conventional ones.
Besides, the 5-year and 30-year come more important than the 10-year maturity itself
(\tref{table-Summary-of-most-influential-relevant-features-weight-higher-than-015}). 
This may be explained by the fact that the 10-year yield is already known to the model.
Third, of note also is the inclusion in those relevant features of 
indicators related to the ECB balance sheet (ECB Balance Sheet Long-Term Refinancing Operations),
at a time when central banks have been involved in large-scale asset purchases or quantitative easing, 
that clearly has an impact on the overall yield levels in the market.
Fourth and last, another example is the 
OECD Leading Indicators in different geographic areas.
These indicators are 
designed with the objective of providing early signals of turning points in economic cycles
and so it is interesting to see them identified as relevant features using this methodology.

%
%
In summary, the LSTM model captures important data and incorporates the information into its long and short-term memories.
Ultimately, the LSTM-LagLasso methodology can also be used for features selection given the richness of information contained in the hidden and cell states.

\subsection{Strength of results}
\label{section-NewLSTMLagLassoMethod-Strength-of-results}

\noindent
In this section, we evaluate the strength of results by assessing whether they could be obtained by chance.
The hypothesis we want to test is whether the results obtained with real features and 
with Gaussian random variables could be part of the same distribution.
For that purpose, we apply the LSTM-LagLasso method 
replacing the macroeconomic and market features by
the same number of Gaussian random variables.
The corresponding mean squared errors are then calculated for each experiment.

\begin{figure}[!htb]
	\centering
	\includegraphics[width=0.475\textwidth]{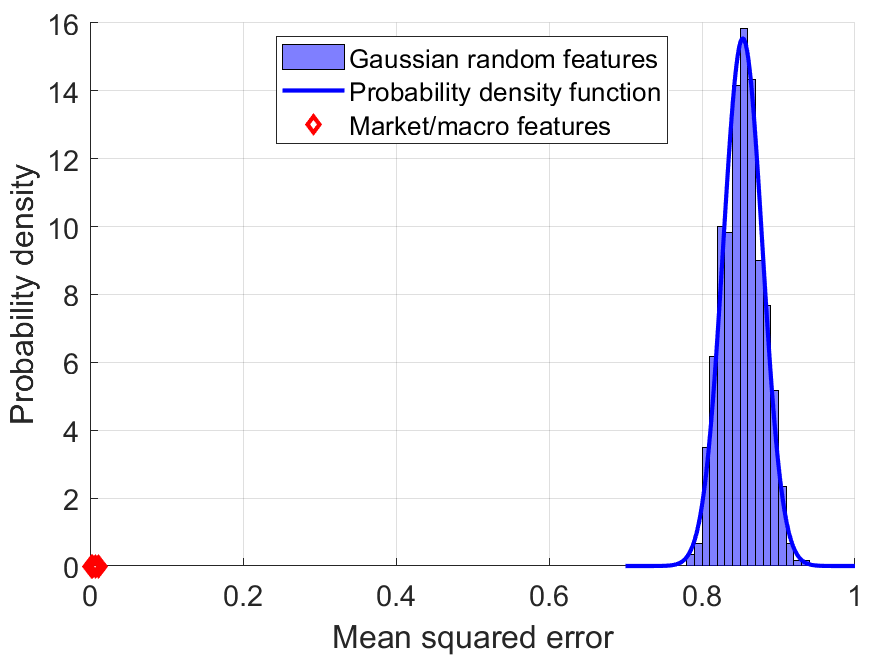}

	\captionsetup{width=0.475\textwidth}
	\caption{Forecasting error of LSTM-LagLasso using macroeconomic / market features and using Gaussian random features.}
	\label{figure-LagLasso_Histogram_Gaussian_Random_Features}
\end{figure}

The simulation is run one hundred times in order to determine the corresponding probability density function and
the results are presented in \fref{figure-LagLasso_Histogram_Gaussian_Random_Features}.
From these we can safely conclude, with statistical significance, that the results with real features are not obtained by chance.

\section{Conclusions and future work}
\label{section-Conclusions-and-future-work}


\noindent
This work has three main components.
%
%
%
First, 
we conduct an application of LSTM networks to the bond market, specifically for forecasting the 10-year Euro government bond yield, 
and compare the results to memory-free standard feedforward neural networks, in particular MLPs. 
This is the first study of its kind as can be confirmed by the lack of published literature in this area.
%
%
To this end, we model the 10-year bond yield using univariate LSTMs with different input sequences (6, 21 and 61 time steps), considering five forecasting horizons, the next day as well as further into the future, up to next day plus 20 days.
Our objective is to compare those LSTM models with univariate MLPs, as well as MLPs using the most relevant features. These are determined using Lasso regression, for each forecasting horizon.
We closely follow the same data and methodology for this comparison. 
In addition, the use of training moving windows incorporating the most recent information as it becomes available has the advantage of increased flexibility to changing market conditions.

The direct comparison of models in identical conditions show that, with the LSTM, we can obtain results that are similar or better and with lower standard deviations. 
In the comparison with the LSTMs using different input sequences, especially for forecasting horizons equal to 10 and 15 days, we observe that the LSTMs with longer input sequences achieve similar levels of forecasting accuracy to the MLP with the most relevant features, with lower standard deviation. 
In other words, the univariate LSTM model with additional memory is capable of achieving similar results as the multivariate MLP with additional information from markets and the economy. 
This is a remarkable achievement and a promising result for future work.
Furthermore, the results for the univariate LSTM show that shorter forecasting horizons require smaller input sequences and, vice-versa.
Therefore, there is a need to adjust the LSTM architecture to the forecasting horizon and in general terms to the conditions of the problem.

In summary, the results obtained in the empirical work validate the potential of LSTMs for yield forecasting and identify their memory advantage when compared to memory-free models.
%
This enables the incorporation of LSTMs in autonomous systems for the asset management industry, with special relevance to pension funds, insurance companies and investment funds.


Second, 
with the objective to analyse the internal functioning of the LSTM model and 
mitigate the preconceived notion of black box normally associated with this type of model,
we conduct an in-depth internal analysis of the information in the memory cell through time.
This is the first contribution with that objective. 
Alternative works are either applied to a different type of model, 
or conduct an external analysis of the LSTMs (\sref{section-LiteratureReview-Main-scope-in-the-financial-applications}).
%
%
To achieve this goal, we select several locations within the memory cell to directly calculate and extract the signals (weights) at each time step and hidden unit.
Specifically, the locations are as follows:
	forget gate,
	product of the outputs from the input gate and input node,
	output gate,
	cell state, and
	hidden state.
This analysis is carried out using 
sequence-to-sequence (6 days) LSTM architectures,
with uni and multivariate feature sets (
10-year yield; 
10-year yield plus momentum indicator, and 
10-year yield plus 5 and 30-year yield),
with reduced number of hidden units (3 units), for interpretability purposes, and
for a forecasting horizon of next day plus 5 days.

Overall, considering all feature sets,
the most remarkable property found consistently in the LSTM signals, 
is the activation / deactivation of units through time,
thus contributing or not (respectively) to the forecasting process. 
Moreover, we found evidence that the LSTM units tend to specialise 
in different yield ranges 
or features considered in the model.

In the third study / contribution, we investigate the information contained in the signals extracted from the LSTM hidden and cell states,
to examine whether the corresponding time series can be explained by external sources of information.
%
%
To this effect, we introduce a new methodology here identified as LSTM-LagLasso, 
based on both Lasso and Kalman-LagLasso.
This methodology is capable of identifying both relevant features and corresponding lags, 
as the Kalman-LagLasso, but with significant modifications
(\sref{section-NewLSTMLagLassoMethod-Methodology} -- LSTM-LagLasso).
%
%
%
%

The findings show that the information contained in the LSTM states is complex, 
but may be explained by exogenous macroeconomic and markets variables, not known to the model during the learning process.
%
Thus, it is worth exploring this information using the developed LSTM-LagLasso methodology, 
which may be used as an alternative feature selection method.
%
On the relevant features selected with the LSTM-LagLasso method,
they indicate conventional as well as non-conventional market/macro indicators (\sref{section-NewLSTMLagLassoMethod-Results-and-discussion}), 
contributing to the prediction process, 
but which are not commonly used in forecasting models.
%
In addition, the LSTM-LagLasso identifies lags as important, in particular $t$, $t-1$ and $t-5$. 
%
Above all, LSTM networks can capture this information and maintain it in the long and short-term memories, i.e. cell and hidden states.

With respect to future work, 
our present research focuses on financial asset forecasting, development of methodologies and analysing internally the LSTM model.
However, the ultimate purpose in the industry is portfolio management and trading. 
In relation to asset forecasting, this is a different type of problem. Obtaining correct predictions does not necessarily translate into profitable strategies. 
Thus, the next step is to implement this type of model in autonomous systems,
to assess its potential for trading and portfolio management in fixed income markets.
Finally, we want to emphasise that the work described in this paper is 
a fundamental component necessary for the implementation of those intelligent systems.




\section*{Acknowledgements}

\noindent 
This work is supported by the UK Engineering and Physical Sciences Research Council (EPSRC Award No. 1921702).
%
All the information required to download the full dataset used in this research (in particular the identification of features), from a Bloomberg Professional terminal, is made publicly available \citep{nunes2019dssPaperDataset}. 
The authors would like to thank Luis Montesdeoca for helpful discussions during the course of this study.


\section*{References}
\Urlmuskip=0mu plus 1mu\relax
\bibliographystyle{apalike}








\end{document}